\documentclass[submission,Phys]{SciPost}
\usepackage{float}
\usepackage{amsmath}
\usepackage{graphicx}
\usepackage{esint}

\makeatletter
\DeclareSymbolFont{usualmathcal}{OMS}{cmsy}{m}{n}
\DeclareSymbolFontAlphabet{\mathcal}{usualmathcal}

\makeatother

\begin{document}

\begin{center}
\textbf{\Large{}Quantum chaos in a harmonic waveguide with scatterers}\\
\textbf{\Large{} }{\Large\par}
\par\end{center}


\begin{center}
Vladimir A. Yurovsky$^{*}$ 
\par\end{center}


\begin{center}
School of Chemistry, Tel Aviv University, 6997801 Tel Aviv, Israel
\\
$^{\star}$\textsf{\small{}volodia@post.tau.ac.il}\textsf{ }
\par\end{center}

\begin{center}
\today 
\par\end{center}


\section*{Abstract}

\textbf{
A set of zero-range scatterers along its axis lifts the integrability
of a harmonic waveguide. Effective solution of the Schr\"odinger
equation for this model is possible due to the separable nature of
the scatterers and millions of eigenstates can be calculated using
modest computational resources. Integrability-chaos transition can
be explored as the model chaoticity increases with the number of scatterers
and their strengths. The regime of complete quantum chaos and eigenstate
thermalization can be approached with 32 scatterers. This is confirmed
by properties of energy spectra, the inverse participation ratio,
and fluctuations of observable expectation values.}

\vspace{10pt}
 \rule{1\textwidth}{1pt} \tableofcontents{}\thispagestyle{fancy}
\rule{1\textwidth}{1pt} \vspace{10pt}

\section{Introduction}

\noindent \label{sec:intro} 
Completely-chaotic systems have impredictable ergodic trajectories
(see \cite{zaslavsky}) and their average properties can be described
by the Gibbs statistical ensemble \cite{landau2013statistical}. In
quantum systems, the statistical description is a consequence of the
eigenstate thermalization hypothesis (ETH), introduced in \cite{deutsch1991,srednicki1994}
(see also \cite{rigol2008,khodja2015}, the experimental work \cite{kaufman2016},
the review \cite{deutsch2018} and the references therein). Energy
spectra of completely-chaotic systems follow Wigner-Dyson statistics
with a dip at small level spacings \cite{guhr1998,mehta,kota}. In
contrast, trajectories of classical integrable systems are completely
predictable and, according to the Kolmogorov-Arnold-Moser theorem,
this property remains even when a weak integrability-breaking perturbation
is applied \cite{zaslavsky}. Quantum systems demonstrate similar
properties (see, e.g., \cite{brandino2015,harshman2017,igloi2023}).
A statistical description by the generalized Gibbs ensemble \cite{rigol2007,doggen_2014,nandy2016,verstraelen2017,wu2019,moller2020}
is applicable to the final state of integrable system relaxation.
The Poisson statistics of integrable system energy spectra has no
dip at small spacings \cite{guhr1998,mehta,kota}.

However, a generic system is not completely chaotic nor integrable
(see examples in \cite{seba1990,seba1991,cheon1996,legrand1997,brown2008,Yurovsky2010,Stone2010,yurovsky2011,Yurovsky2011b,kollar2011,Olshanii2012,neuenhahn2012,canovi2012,larson2013,venuti2013,mumford2014,fialko2014,marchukov2014,andraschko2014,khripkov2015,venuti2015,khripkov2016,bartsch2017,dag2018,igloi2018,yesha2018,goldfriend2019,bastianello2019,huang2022,ma2022,sierant2022,pecci2022}).
Certain incompletely-chaotic systems --- the systems with no selection
rules --- relax to a state whose properties are governed by the inverse
participation ratio (IPR) \cite{yurovsky2011,Olshanii2012}. Inverse
of this parameter estimates the number of integrable system eigenstates
comprising the non-integrable one. IPR ranges from 0 for completely-chaotic
systems to 1 for integrable ones. Then it can serve as a measure of
the system's chaoticity \cite{georgeot1997}. IPR also governs fluctuations
of eigenstate expectation values \cite{neuenhahn2012}. The energy-spectrum
statistics of incompletely-chaotic systems lie between the Wigner-Dyson
and Poisson ones. Certain systems demonstrate the \v{S}eba statistics
\cite{seba1991}.

The most obvious objects of chaotic property simulation are lattice
systems. However, they have a finite Hilbert space and its dimension
is restricted due to computational difficulties (complexity of lattice
system simulations increases as a high power of the lattice site number
and exponentially with the number of particles). Then, on increase
of the system chaoticity, each eigenstate can fill the full Hilbert
space. A system with infinite Hilbert space --- the Sinai type billiard
--- was analyzed in \cite{barnett2006}, where $\sim3\times10^{5}$
eigenstates were calculated. However, chaoticity of such billiard
cannot be tuned.

The present model --- a particle in a harmonic waveguide with zero-range
scatterers along its axis --- has an infinite Hilbert space. As the
scatterers are a particular case of independent perturbations \cite{yurovsky2023},
IPR should be inversely proportional to the number of scatterers.
The model chaoticity can also be tuned by the scatterer strengths.
This model was already used in \cite{yurovsky2023} for numerical
confirmation of the general relations between properties of wavefunctions
and the number of scatterers. The present paper is devoted exclusively
to the harmonic waveguide with scatterers and analyzes properties
of wavefunctions for weak perturbations and for additional models,
as well as properties of energy spectra.

Since a zero-range scatterer is a particular case of separable interactions,
the present model belongs to systems with high-rank separable perturbations
\cite{albeverio2000}. Energy spectra of several physical systems
of such type have already been considered. They are the flat rectangular
billiards --- generalization of the \v{S}eba billiard \cite{seba1990}
--- with 1-3 \cite{cheon1996}, 6 \cite{legrand1997}, and 2 \cite{yesha2018}
scatterers. Theoretical predictions for a single scatterer in a harmonic
potential were compared to experiments \cite{guan2019}. Series of
separable interactions can also approximate the dipole-dipole ones
\cite{derevianko2003,derevianko2005}. Energy spectra of two dipolar
particles in a harmonic trap were calculated \cite{kanjilal2007}
using such expansion. An advantage of systems with separable rank-$s$
interactions is that calculations require diagonalization of a $s\times s$
matrix, (cf. to $\alpha\times\alpha$ matrix in the direct diagonalization
method for $\alpha$ eigenstates). In addition, the present model
allows an analytical summation over axial states. Then the system
properties are calculated here for millions of eigenstates.

The paper has the following organization. The model is described in
Sec. \eqref{sec:The-model}. Section \eqref{sec:Energy-spectra} analyzes
the energy spectra statistics. Properties of wavefunctions, including
expectation value fluctuations and IPR, are presented in Sec. \ref{sec:PropWaveFun}.
Appendices provide derivation details and additional technical information. 

A system of units in which Planck\textquoteright s constant is $\hbar=1$
is used below. 

\section{The model\label{sec:The-model}}

The Hamiltonian of a particle with the mass $m$ in an axially-symmetric
harmonic waveguide with the transverse frequency $\omega_{\perp}$
contains the kinetic and potential energies, 
\begin{equation}
\hat{H}_{0}=\frac{1}{2m}\left[\left(\frac{1}{i}\frac{\partial}{\partial z}-A\right)^{2}-\triangle_{\rho}\right]+\frac{m\omega_{\perp}^{2}\rho^{2}}{2}.\label{eq:H0}
\end{equation}
Here $z$ is the axial coordinate, $\rho=\sqrt{x^{2}+y^{2}}$ is the
transverse radius, $\triangle_{\rho}$ is the transverse Laplacian,
and $A$ is the vector potential (its role will be discussed below).

Integrability of the perturbed Hamiltonian 
\begin{equation}
\hat{H}_{s}=\hat{H}_{0}+\sum_{s'=1}^{s}\hat{V}_{s'}
\end{equation}
is lifted by the zero-range scatterers
\begin{equation}
\hat{V}_{s'}=V_{s'}\delta_{reg}(\mathbf{r}-\mathbf{R}_{s'}),\label{eq:Vs}
\end{equation}
where $\delta_{reg}$ is the Fermi-Huang pseudopotential and the scatterers
are located along the waveguide axis, i.e., their positions $\mathbf{R}_{s'}=(0,0,z_{s'})$
have zero transverse coordinates. The scatterers are numbered from
left to right ($z_{s'}>z_{s''}$ if $s'>s''$). The model is restricted
in the sector of the axially-symmetric states, as other states vanish
at the waveguide axis, and, therefore, are not affected by the scatterers.
Then the eigenstates of $\hat{H}_{0}$, labeled by the axial $l$
and radial $n\geq0$ quantum numbers, are $\left\langle \rho,z|nl\right\rangle =\left\langle \rho|n\right\rangle \left\langle z|l\right\rangle $
with the radial wavefunctions
\begin{equation}
\left\langle \rho|n\right\rangle =\frac{1}{\sqrt{\pi}a_{\perp}}L_{n}^{(0)}((\rho/a_{\perp})^{2})\exp(-(\rho/a_{\perp})^{2}/2).\label{eq:rhon}
\end{equation}
Here $a_{\perp}=(m\omega_{\perp})^{-1/2}$ is the transverse oscillator
range and $L_{n}^{(0)}$ are the Laguerrre polynomials (see \cite{DLMF}).
The discrete energy spectrum is provided either by the periodic boundary
conditions (PBC) $\left\langle z+L|l\right\rangle =\left\langle z|l\right\rangle $,
or by the hard-wall box (HWB) $\left\langle z=L|l\right\rangle =\left\langle z=0|l\right\rangle =0$.
Then the axial wavefunctions are either
\begin{equation}
\left\langle z|l\right\rangle =L^{-1/2}e^{2i\pi l\zeta}\label{eq:zlpbc}
\end{equation}
with $-\infty<l<\infty$ and $\zeta=z/L$ for PBC or
\begin{equation}
\left\langle z|l\right\rangle =(2/L)^{1/2}\sin\pi l\zeta\label{eq:zlhwb}
\end{equation}
with $1\leq l<\infty$ for HWB.

The particular case of PBC with a single scatterer was considered
in \cite{Yurovsky2010,Stone2010,yurovsky2011,Yurovsky2011b}.

For PBC, the eigenstate $\left|nl\right\rangle $ of $\hat{H}_{0}$
has the eigenenergy
\begin{equation}
E_{nl}=\frac{2}{mL^{2}}\varepsilon_{nl}+\omega_{\perp},\quad\varepsilon_{nl}=\lambda n+\pi^{2}(l-l_{0})^{2},\label{eq:Enl}
\end{equation}
where $\lambda=(L/a_{\perp})^{2}$ characterizes the aspect ratio
and $l_{0}=LA/(2\pi)$ is the scaled vector potential. If $A=0$,
the inversion (P) invariance of the Hamiltonian $\hat{H}_{0}$ leads
to the degeneracy of the energies $E_{nl}$ and $E_{n-l}$ . This
degeneracy can be lifted by any P-noninvariant perturbation. The vector
potential lifts it as well, with no effect on the simple wavefunctions
\eqref{eq:zlpbc}, though the Hamiltonian losses the time-reversal
(T) invariance. 

Four kinds of the model are considered here. The first three kinds
correspond to PBC. The first, non-symmetric, model has $A\neq0$ and
is T-noninvariant. The scatterer positions 
\begin{equation}
z_{1}=0,\quad z_{s'}=(s'-1+\delta_{s'})L/s\quad(s'>1)\label{eq:zsnosymm}
\end{equation}
form irregular sequence due to random shifts $-0.25\leq\delta_{s'}<0.25$.
The shifts are calculated once for each number of scatterers and there
is no average over the shifts. In the second, symmetric, model with
$z_{s-s'+1}=z_{s}-z_{s'}+z_{1}$ for $s'>s/2$, the scatterer positions
are invariant over inversion under $(z_{1}+z_{s})/2$. This inversion
changes the sign of the term $(i/m)A\partial/\partial z$ in the Hamiltonian
$\hat{H}_{0}$. This sign is also changed by the time-reversal (complex
conjugation). Then the symmetric model with equal $V_{s'}$ is PT-invariant.
The third, T-invariant, model has $A=0$ and the same scatterer positions
as the non-symmetric one. Only this model has a degenerate energy
spectrum of the integrable Hamiltonian. The fourth, box, model corresponds
to HWB. The scatterer positions are $z_{s'}=(s'+\delta_{s'})L/(s+1)$.
Although $A=0$, the energy spectrum 
\begin{equation}
\varepsilon_{nl}=\lambda n+\frac{\pi^{2}}{4}l^{2}\label{eq:enlhwb}
\end{equation}
is non-degenerate as $l$ is positive.

Together, the four kinds of the model cover different symmetries of
the Hamiltonian (T-invariant, PT-invariant, and non-symmetric), as
well as different boundary conditions (PBC and HWB). 

The eigenstates of the non-integrable system $\left|\alpha\right\rangle $,
solutions to the Schr\"{o}dinger equation $\hat{H}\left|\alpha\right\rangle =E_{\alpha}\left|\alpha\right\rangle $,
are labeled in the increasing order of the eigenenergies $E_{\alpha}$.
Expansion over the integrable system eigenstates $\left|nl\right\rangle $
transforms the Schr\"{o}dinger equation to the form
\begin{equation}
\left|\alpha\right\rangle =\sum_{n,l}\frac{\left|nl\right\rangle \left\langle nl\right|}{E_{\alpha}-E_{nl}}\sum_{s'=1}^{s}\hat{V}_{s'}\left|\alpha\right\rangle .\label{eq:LippShw}
\end{equation}
According to \eqref{eq:Vs}
\begin{equation}
\left\langle nl\left|\hat{V}_{s'}\right|\alpha\right\rangle =V_{s'}\left\langle nl|\mathbf{R}_{s'}\right\rangle \left\langle \mathbf{R}_{s'}|\alpha\right\rangle _{reg},\label{eq:nlVsalpha}
\end{equation}
where the value of the regular part of $\left|\alpha\right\rangle $
at $\mathbf{R}_{s'}$ is 
\begin{equation}
\left\langle \mathbf{R}_{s'}|\alpha\right\rangle _{reg}=\frac{\partial}{\partial r}\left[r\left\langle \mathbf{r}|\alpha\right\rangle \right]_{\mathbf{r}=\mathbf{R}_{s'}}=\frac{\partial}{\partial z}\left[z\left\langle 0,0,z|\alpha\right\rangle \right]_{z=z_{s'}}.
\end{equation}
The last equality above follows from the spherical symmetry of $\left\langle \mathbf{r}|\alpha\right\rangle $
in the vicinity of $\mathbf{R}_{s'}$ \cite{moore2004}. As a result,
we get the following system of linear equations for $\left\langle \mathbf{R}_{s'}|\alpha\right\rangle _{reg}$
\begin{equation}
\left\langle \mathbf{R}_{s'}|\alpha\right\rangle _{reg}=\sum_{s''=1}^{s}V_{s''}\frac{\partial}{\partial z}\left[z\sum_{n,l}\frac{\left\langle 0,0,z|nl\right\rangle \left\langle nl|0,0,z_{s''}\right\rangle }{E_{\alpha}-E_{nl}}\right]_{z=z_{s'}}\left\langle \mathbf{R}_{s''}|\alpha\right\rangle _{reg}.\label{eq:linsys}
\end{equation}
For the wavefunctions \eqref{eq:rhon} and \eqref{eq:zlpbc} or \eqref{eq:zlhwb}
and energies \eqref{eq:Enl} or \eqref{eq:enlhwb} the sum over $l$
above can be calculated analytically (see Appendix \ref{sec:DerivationT}).
Then the system \eqref{eq:linsys} attains the form
\begin{equation}
\sum_{s''=1}^{s}S_{s's''}(\varepsilon)\left\langle \mathbf{R}_{s''}|\alpha\right\rangle _{reg}=0\label{eq:linsysreg}
\end{equation}
with
\begin{align}
S_{s's''}(\varepsilon) & =\frac{V_{s''}}{V_{0}}\sqrt{\lambda}\sum_{n=0}^{\infty}T_{n}(\zeta_{s'},\zeta_{s''})\quad(s'>s''),\qquad S_{s''s'}(\varepsilon)=S_{s's''}^{*}(\varepsilon)\label{eq:Sspspp}\\
S_{s's'}(\varepsilon) & =\frac{V_{s'}}{V_{0}}\left[\sqrt{\lambda}\left(\sum_{n=0}^{[\varepsilon_{\alpha}/\lambda]}T_{n}(\zeta_{s'},\zeta_{s'})+\sum_{n=[\varepsilon_{\alpha}/\lambda]+1}^{\infty}T_{n}^{reg}(\zeta_{s'})\right)-\zeta\left(\frac{1}{2},\left[\frac{\varepsilon}{\lambda}\right]+1-\frac{\varepsilon}{\lambda}\right)\right]-1.\label{eq:Sspsp}
\end{align}
Here $[]$ denote the integer part, $\zeta_{s'}=z_{s'}/L$, $\zeta(.,.)$
is the Hurwitz zeta function (see \cite{DLMF}), $V_{0}=2\pi a_{\perp}/m$
is the scale of the interaction strength, and the summands $T_{n}(\zeta_{s'},\zeta_{s''})$
and $T_{n}^{reg}$ are given in Appendix \ref{sec:DerivationT} for
each kind of the model. Due to arrangement of scatterers, only $T_{n}(\zeta_{s'},\zeta_{s''})$
with $\zeta_{s'}\geq\zeta_{s''}$ have to be calculated. $T_{n}(\zeta_{s'},\zeta_{s'})$
and $T_{n}^{reg}$ are always real functions. If $A=0$, $T_{n}(\zeta_{s'},\zeta_{s''})$,
as well as the matrix $S_{s's''}(\varepsilon)$, is real, and $S_{s's''}(\varepsilon)$
is symmetric.

The system \eqref{eq:linsysreg} has a non-trivial solution at $\varepsilon=\varepsilon_{\alpha}\equiv mL^{2}(E_{\alpha}-\omega_{\perp})/2$
where an eigenvalue of its matrix has a root as a function of $\varepsilon$.
The matrix $S_{s's''}(\varepsilon)$ has poles at $\varepsilon=\varepsilon_{nl}$,
as it is seen from \eqref{eq:linsys}. Then the eigenvalues can have
poles at $\varepsilon=\varepsilon_{nl}$ as well. Between these poles
each eigenvalue is a monotonic function of $\varepsilon$, as demonstrated
by direct calculations (see Appendix \ref{sec:Eigenvalues}). Than
all eigenenergies $\varepsilon_{\alpha}$ in each interval between
neighboring $\varepsilon_{nl}$ can be calculated as roots of $s$
eigenvalues. Although the eigenvalue monotonicity was not proved exactly,
this algorithm provides the number of eigenenergies $\varepsilon_{\alpha}$
which differs from the number of $\varepsilon_{nl}$ in the same energy
interval by not more than $s$. It is an evidence that no eigenenergies
$\varepsilon_{\alpha}$ are lost.

The terms in the sums over $n$ in Eqs. \eqref{eq:Sspspp} and \eqref{eq:Sspsp}
decay exponentially when $n>\varepsilon_{\alpha}/\lambda$ (see Appendix
\ref{sec:DerivationT}). Thus, the calculation of the system \eqref{eq:linsysreg}
matrix requires $\propto s^{2}\alpha^{2/3}$ operations since $\varepsilon_{\alpha}\propto\alpha^{2/3}$
{[}see Eq. \eqref{eq:alphabarPBC} below{]}, while its solution requires
$s^{3}$ operations. Then, if $s\ll\alpha^{2/3}$, calculation of
$\alpha$ eigenenergies requires $\propto s^{2}\alpha^{5/3}$ operations
--- much less than $\alpha^{3}$ operations in the direct diagonalization
method.

There seems to be no fundamental obstacle for experimental realization
of the present model. In the case of cold trapped atoms, atoms of
other kind in optical tweezers might play the role of scatterers,
and the interaction strength might be tuned by a Feshbach resonance.
T-noninvariant models might be realized with trapped ions in a magnetic
field. In optics, optical defects might work as scatterers \cite{bruck2016}
for photons in an optical cavity or waveguide (see also \cite{valagiannopoulos2007,mandilara2019}
and the references therein). The PBC models might be realized with
circular atomic or optical waveguides.

\section{Statistics of energy spectra \label{sec:Energy-spectra}}

The differences in energy spectra between integrable and chaotic systems
were the first distinctive properties of quantum chaos (see \cite{guhr1998,mehta,kota}).
These properties are defined in terms of the unfolded energy $\bar{\alpha}(\varepsilon_{\alpha})$
--- the smooth part of the dependence $\alpha(\varepsilon_{\alpha})$.
For the present model, the unfolding function is the same as for the
underlying integrable system. The number of states below the scaled
energy $\varepsilon$ is the staircase function
\begin{equation}
\alpha(\varepsilon)=\sum_{n,l}\theta(\varepsilon-\varepsilon_{nl}).
\end{equation}
For PBC, using Eq. \eqref{eq:Enl} for $\varepsilon_{nl}$, we have
\begin{equation}
\alpha(\varepsilon)=\sum_{l=-\infty}^{\infty}\left\{ [(\varepsilon-\pi^{2}(l-l_{0})^{2})/\lambda]+1\right\} \theta((\varepsilon-\pi^{2}(l-l_{0})^{2})/\lambda+1).
\end{equation}
The smooth part is extracted by replacing the integer part $[x]$
with $x-1/2$. The limits of the sum over $l$, $[l_{0}\pm\sqrt{\varepsilon}/\pi]$,
are replaced in the same way. As a result, we get
\begin{equation}
\bar{\alpha}(\varepsilon)=\frac{4}{3\pi\lambda}\varepsilon^{3/2}+\left(\frac{1}{\pi}+\frac{\pi}{6\lambda}\right)\varepsilon^{1/2}.\label{eq:alphabarPBC}
\end{equation}
The $\varepsilon$-independent terms are dropped here, since only
differences between $\bar{\alpha}(\varepsilon_{\alpha})$ appear in
the following expressions. Similar expression is obtained for the
HWB model 
\begin{equation}
\bar{\alpha}(\varepsilon)=\frac{4}{3\pi\lambda}\varepsilon^{3/2}-\frac{\varepsilon}{2\lambda}+\left(\frac{1}{\pi}+\frac{\pi}{24\lambda}\right)\varepsilon^{1/2}.
\end{equation}

The first property of the energy spectrum considered here is the nearest-neighbor
distribution (NND) --- the density of probability to have the given
value of the unfolded energy difference $\bar{\alpha}=\bar{\alpha}(\varepsilon_{\alpha})-\bar{\alpha}(\varepsilon_{\alpha-1})$
between the neighboring energy levels \cite{guhr1998,mehta,kota}.
Integrable systems have the Poisson NND,
\begin{equation}
w_{Pois}(\bar{\alpha})=e^{-\bar{\alpha}},\label{eq:wPois}
\end{equation}
while completely-chaotic ones have the Wigner-Dyson distributions
for Gaussian ensembles of random orthogonal matrices (GOE) 
\begin{equation}
w_{GOE}(\bar{\alpha})=\frac{\pi}{2}\bar{\alpha}\exp\left(-\frac{\pi}{4}\bar{\alpha}^{2}\right)\label{eq:wGOE}
\end{equation}
and unitary matrices (GUE)
\begin{equation}
w_{GUE}(\bar{\alpha})=\frac{32}{\pi^{2}}\bar{\alpha}^{2}\exp\left(-\frac{4}{\pi}\bar{\alpha}^{2}\right)\label{eq:wGUE}
\end{equation}
in the cases of T-invariant and T-noninvariant systems, respectively.

The \v{S}eba NND 
\begin{equation}
w_{Seba}(\bar{\alpha})=A_{Seba}\bar{\alpha}\exp\left(-B_{Seba}\bar{\alpha}-\frac{A_{Seba}}{B_{Seba}^{2}}\left[1-e^{-B_{Seba}\bar{\alpha}}(B_{Seba}\bar{\alpha}+1)\right]\right)\label{eq:wSeba}
\end{equation}
with $A_{Seba}\approx2.1266$ and $B_{Seba}\approx0.3481$ was obtained
\cite{seba1991} for certain incompletely-chaotic systems.

All states of a non-integrable system correspond to the same symmetry
and then their energies demonstrate repulsion. Then NND \eqref{eq:wGOE},\eqref{eq:wGUE},
and \eqref{eq:wSeba} of non-integrable systems vanish at the zero
level spacing and decrease approaching this point. The integrable
system states of different symmetry can be energy degenerate, and
then NND \eqref{eq:wPois} decreases exponentially with the level
spacing. For non-integrable systems NND decreases at large spacing
too, although completely-chaotic systems are characterized by Gaussian
decrease {[}see Eqs. \eqref{eq:wGOE} and \eqref{eq:wGUE}{]}, while
the \v{S}eba NND \eqref{eq:wSeba} decreases exponentially.

Another property of energy spectra is the spectral rigidity $\Delta_{3}(\Delta\bar{\alpha})$
--- the least-square deviation of the staircase function $\alpha(\varepsilon)$
from the best fit to a straight line on a given interval of the unfolded
energy $\Delta\bar{\alpha}$. The spectral rigidity for integrable
and completely-chaotic (T-invariant and T-noninvariant) systems are
given, respectively, by \cite{guhr1998,mehta,kota}
\begin{align}
\Delta_{3}^{Pois}(\Delta\bar{\alpha}) & =\Delta\bar{\alpha}/15\nonumber \\
\Delta_{3}^{GOE}(\Delta\bar{\alpha}) & =\frac{1}{\pi^{2}}\left(\ln\Delta\bar{\alpha}+\ln2\pi+\gamma_{Eul}-\frac{5}{4}-\frac{\pi^{2}}{8}\right)\\
\Delta_{3}^{GUE}(\Delta\bar{\alpha}) & =\frac{1}{2\pi^{2}}\left(\ln\Delta\bar{\alpha}+\ln2\pi+\gamma_{Eul}-\frac{5}{4}\right),\nonumber 
\end{align}
where $\gamma_{Eul}\approx0.5772$ is the Euler's constant \cite{DLMF}.

The energy spectrum properties are calculated below for the four kinds
of the models. The parameters $l_{0}=0.25-e^{-4}\approx0.232$ (for
the T-noninvariant models) and $\lambda=\pi^{3}(1+\sqrt{5})\approx100$
are expressed in terms of transcendent numbers {[}$(1+\sqrt{5})/2$
is the golden ratio{]}. Most of the results are obtained for $10^{6}$
eigenstates in the unitary regime, $V_{s'}=10^{6}V_{0}$ for all scatterers.
\begin{figure}[H]
\includegraphics[width=8cm]{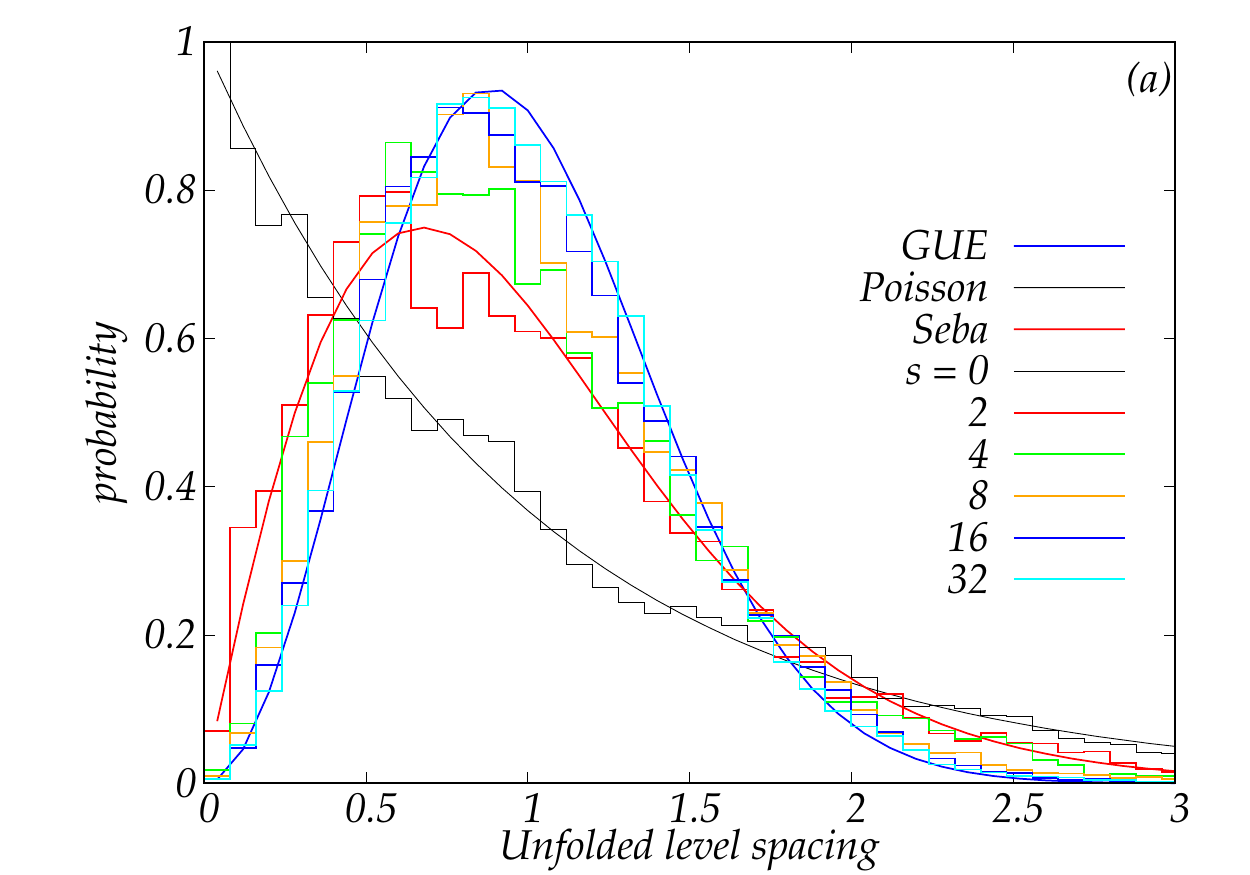}\includegraphics[width=8cm]{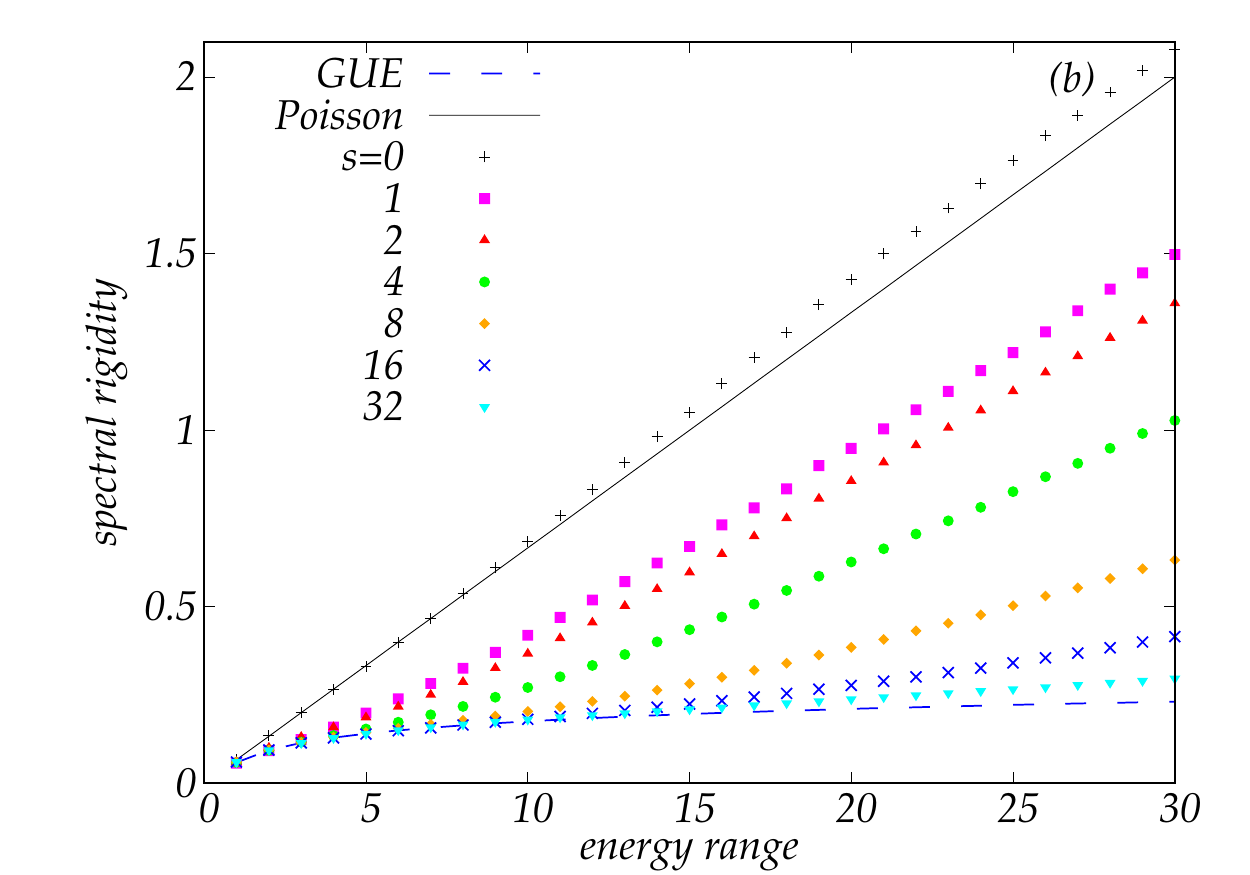}

\caption{Near-neighbor distribution (a) and the spectral rigidity (b) for the
non-symmetric model with different numbers of scatterers in the unitary
regime.\label{fig:NNDSpRigNonSym}}

\end{figure}

Figure \ref{fig:NNDSpRigNonSym}(a) shows NND calculated for the non-symmetric
model with different numbers of scatterers in the unitary regime.
For $s=2$, NND follows the \v{S}eba plot, as well as for the case
of $s=1$ considered in \cite{Stone2010}. When the number of scatterers
increases, NND tends to the GUE prediction and approaches it at $s=32$.
GUE is approached as the model is T-noninvariant. The calculated spectral
rigidity {[}see Fig. \ref{fig:NNDSpRigNonSym}(b){]} demonstrates
the same tendency.
\begin{figure}[H]
\includegraphics[width=8cm]{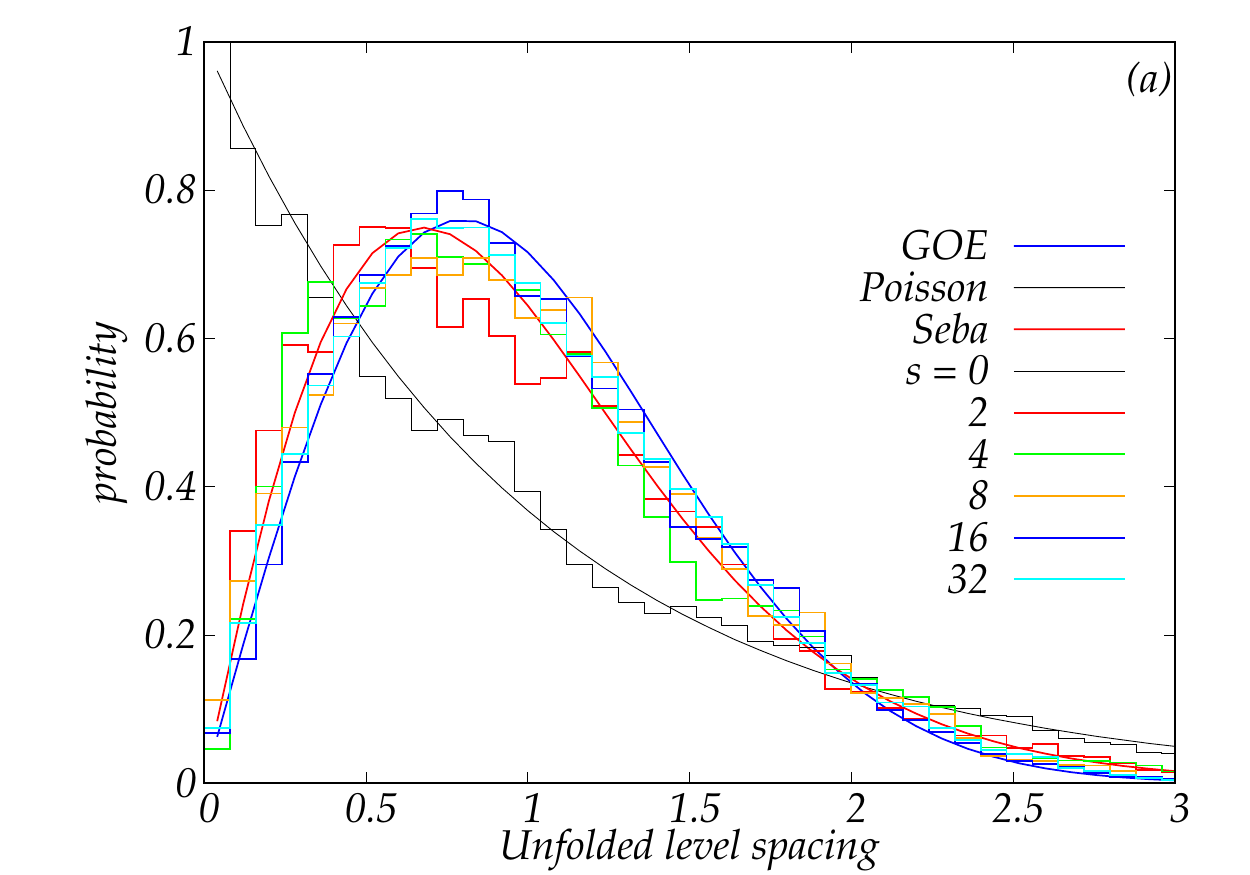}\includegraphics[width=8cm]{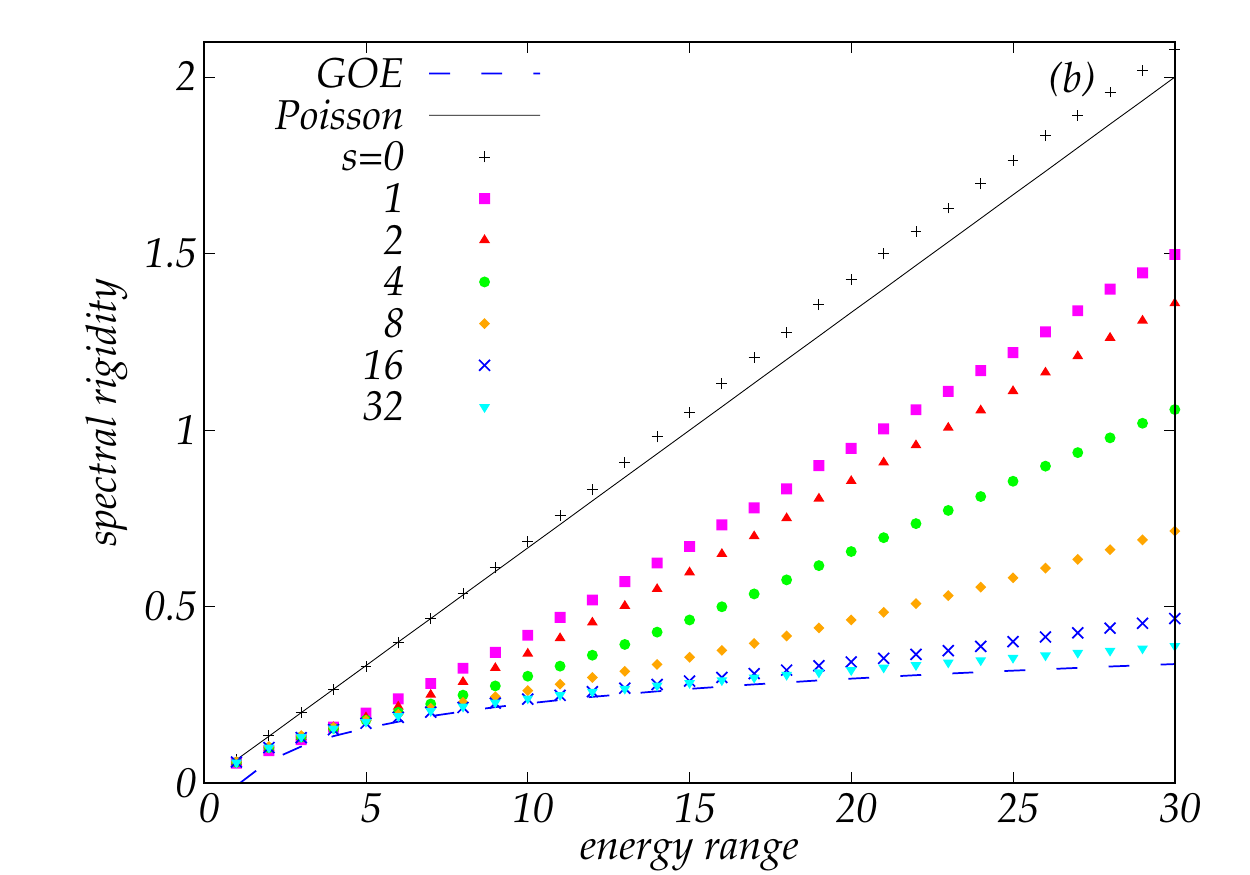}

\caption{Near-neighbor distribution (a) and the spectral rigidity (b) for the
symmetric model with different numbers of scatterers in the unitary
regime.\label{fig:NNDSpRigSym}}

\end{figure}

For the symmetric model (see Fig. \ref{fig:NNDSpRigSym}), NND for
$s=2$ again follows the \v{S}eba predictions (indeed, the case of
two scatterers of the same strength is always P-invariant). However,
at $s=32$, NND and spectral rigidity approach the GOE predictions,
although the system is T-noninvariant. It is a consequence of the
real matrix of the interaction with scatterers 
\begin{equation}
\left\langle n'l'\left|\sum_{s'=1}^{s}\hat{V}_{s'}\right|nl\right\rangle =\frac{V_{1}}{\pi La_{\perp}^{2}}\sum_{s'=1}^{s}\cos2\pi(l-l')\tilde{\zeta}_{s'}
\end{equation}
obtained when the $z$ coordinate origin is shifted to $(z_{1}+z_{s})/2$,
such that $\tilde{\zeta}_{s'}=\zeta_{s'}-(\zeta_{1}+\zeta_{s})/2$.
The real matrix should be described by GOE, like in T-invariant systems.
\begin{figure}[H]
\includegraphics[width=8cm]{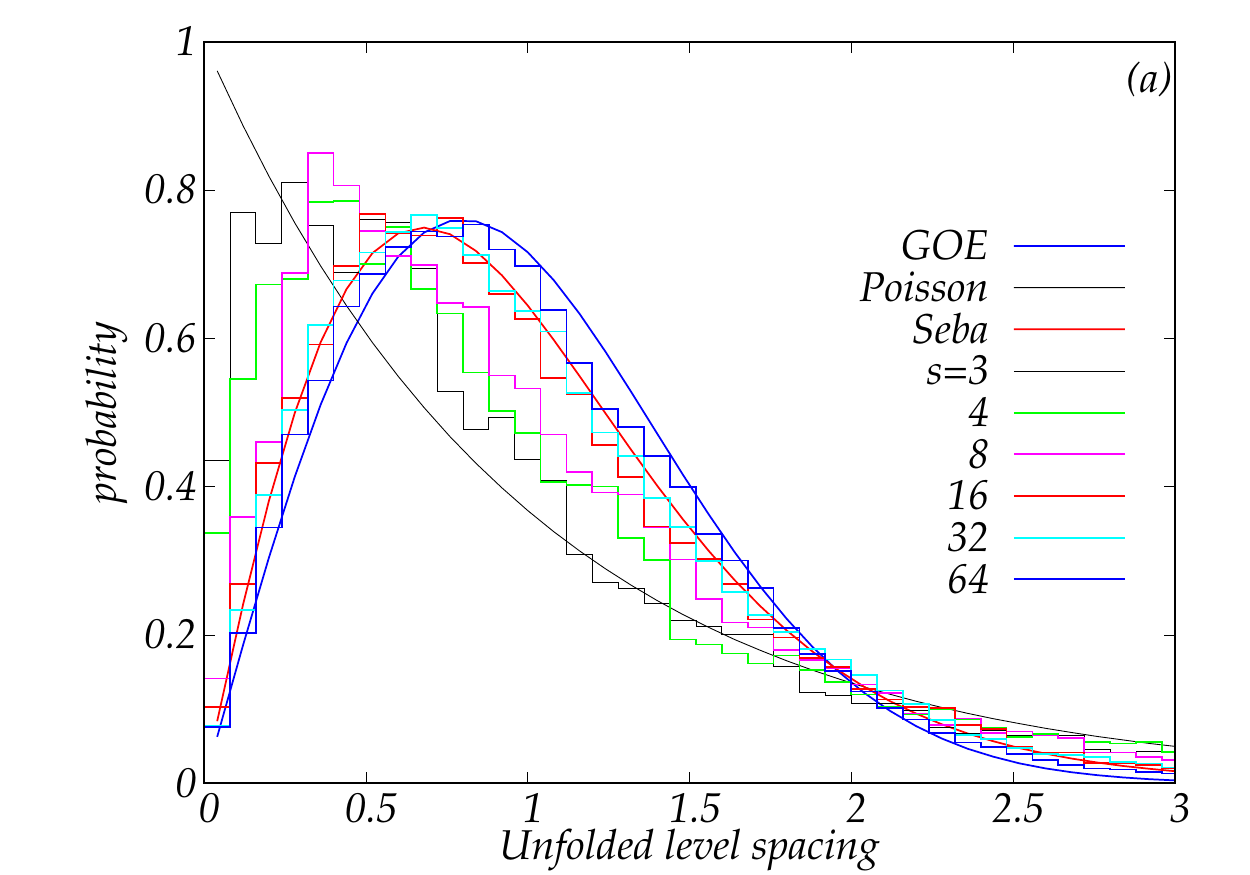}\includegraphics[width=8cm]{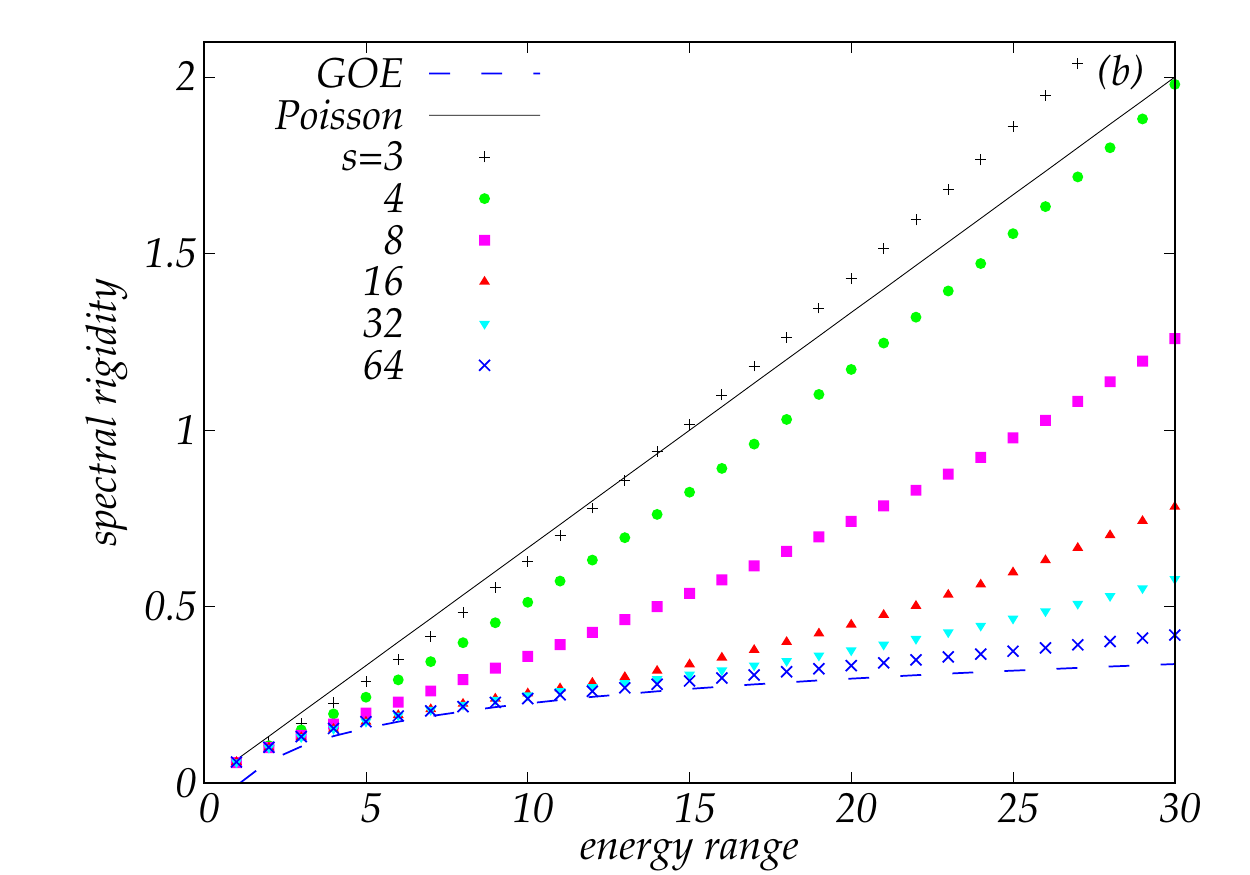}

\caption{Near-neighbor distribution (a) and the spectral rigidity (b) for the
T-invariant PBC model with different numbers of scatterers in the
unitary regime.\label{fig:NNDSpRigT}}

\end{figure}

NND and spectral rigidity for the T-invariant model are shown in Fig.
\ref{fig:NNDSpRigT}. Now \v{S}eba and GOE NND are approached only
at $s=32$ and $s=64$, respectively. For $s=3$ both NND and spectral
rigidity are close to the Poisson predictions. Then, this model is
less chaotic than the T-noninvariant ones where the \v{S}eba and
Wigner-Dyson statistics are approached at $s=2$ and $s=32$, respectively.
This may be related to degeneracy of the integrable system energy
spectrum for the T-invariant model. 
\begin{figure}[H]
\includegraphics[width=8cm]{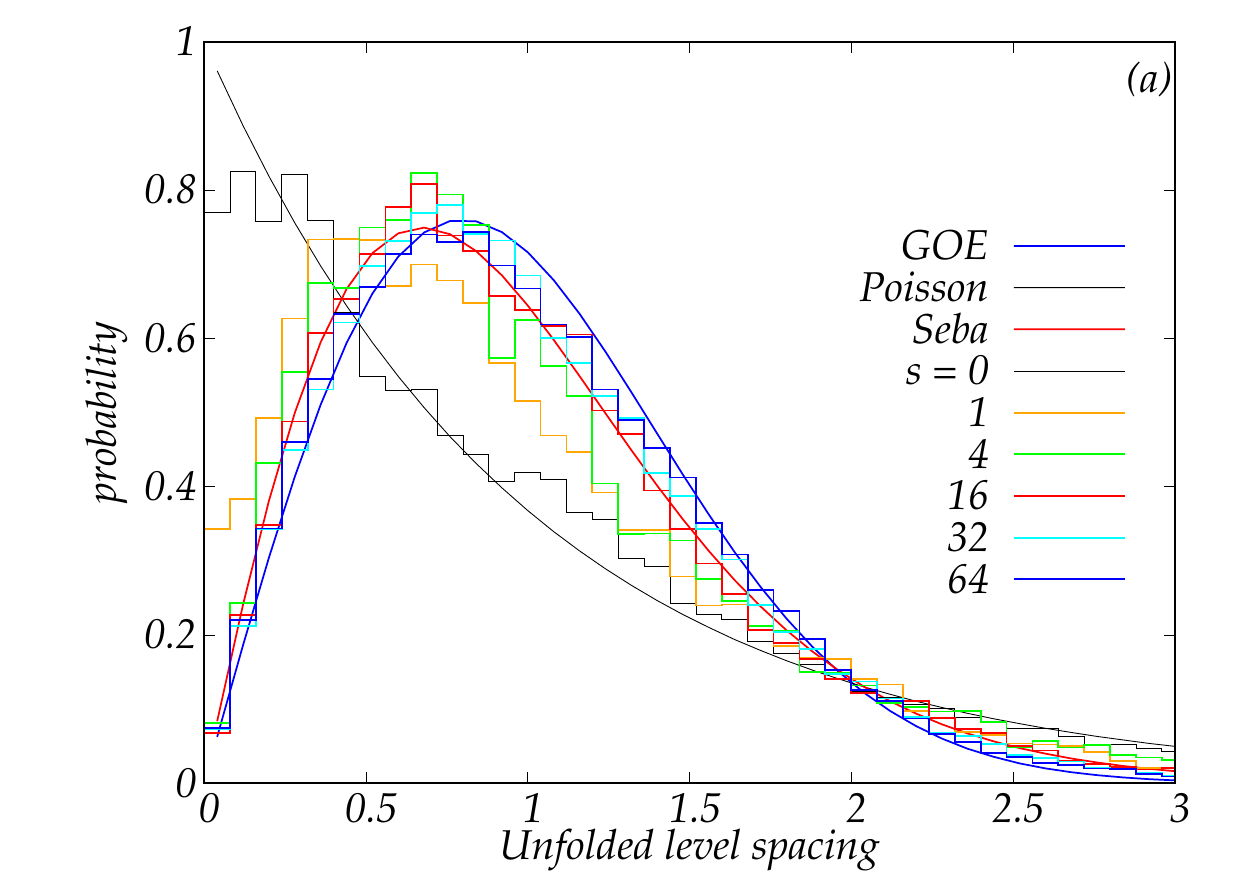}\includegraphics[width=8cm]{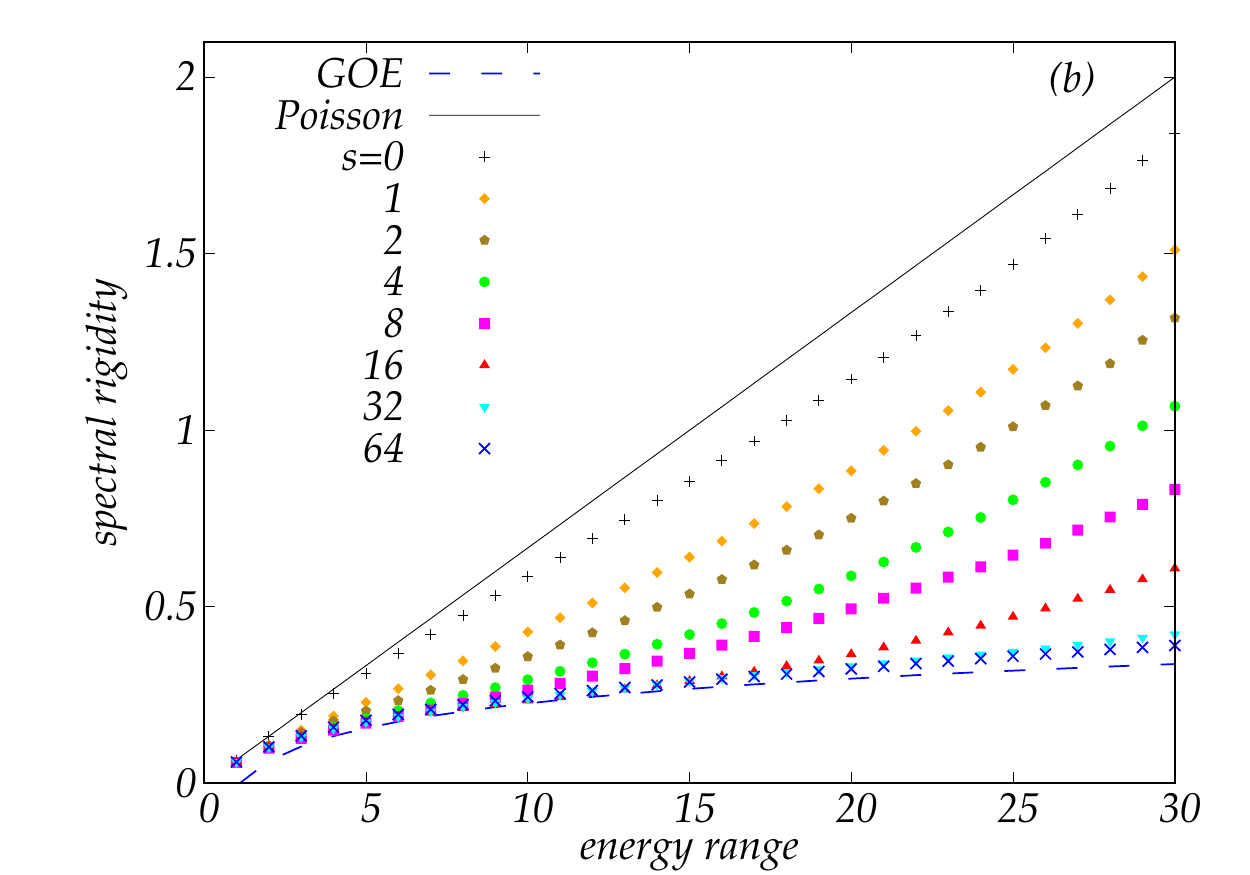}

\caption{Near-neighbor distribution (a) and the spectral rigidity (b) for the
HWB model with different numbers of scatterers in the unitary regime.\label{fig:NNDSpRigB}}

\end{figure}

This assumption is confirmed by the NND and spectral rigidity for
the HWB model (see Fig. \ref{fig:NNDSpRigB}). This model with non-degenerate
energy spectrum is more chaotic than the PBC T-invariant one, as now
\v{S}eba and GOE predictions are approaching at $s=16$ and $s=32$,
respectively. Then, this model is less chaotic than the T-noninvariant
PBC ones. There is also a noticeable difference between these models
in the statistics of integrable system energy spectra --- for the
HWB model NND at small spacings and spectral rigidity are below the
Poisson predictions. 
\begin{figure}[H]
\includegraphics[width=8cm]{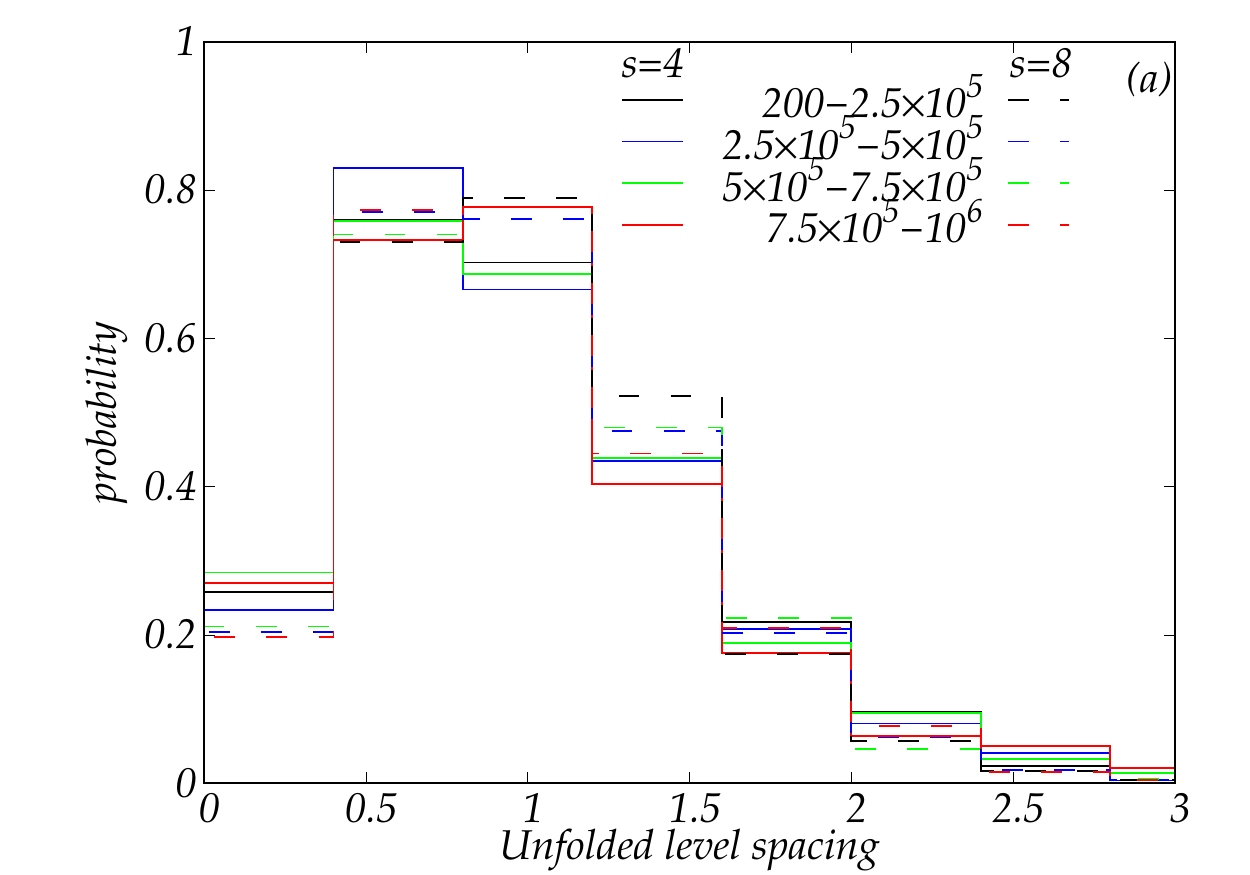}\includegraphics[width=8cm]{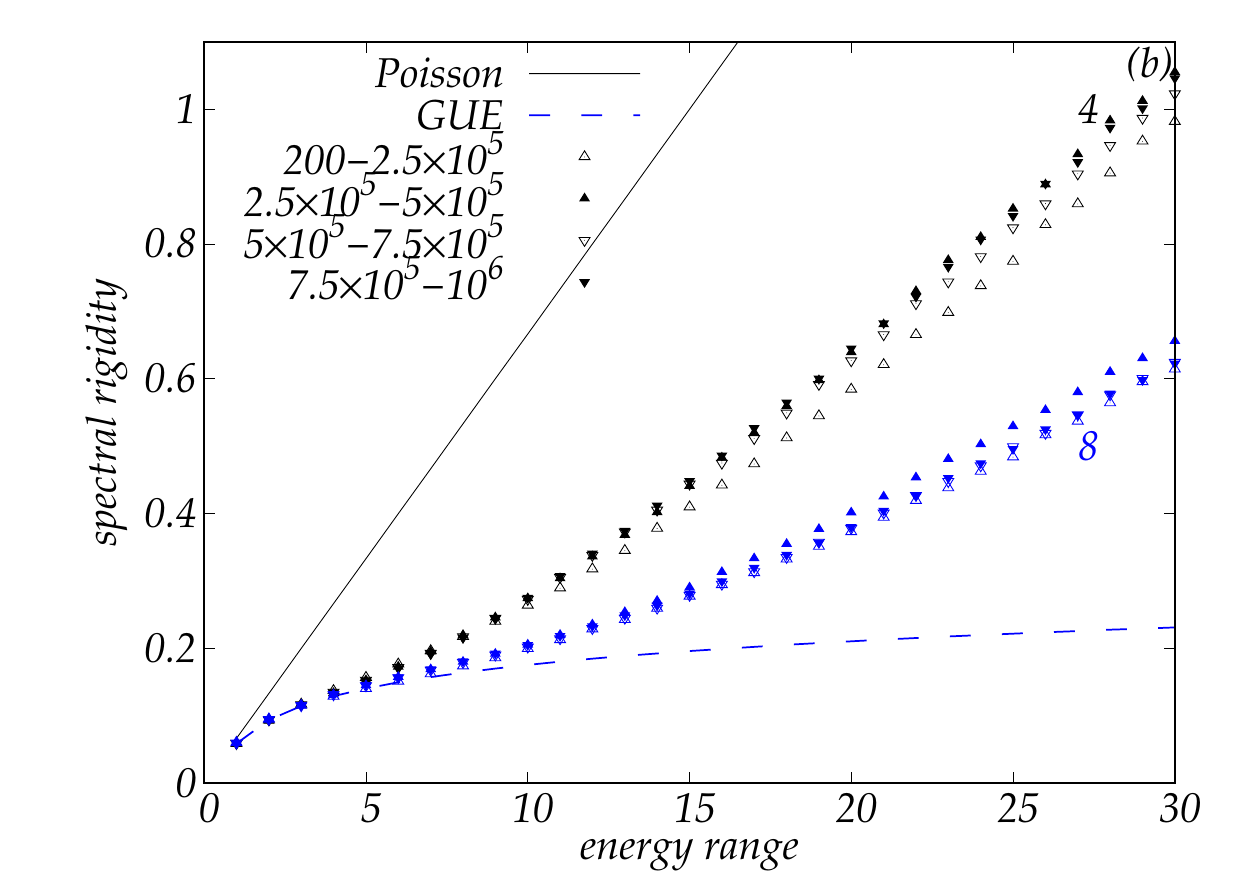}

\caption{Near-neighbor distribution (a) and the spectral rigidity (b) for the
non-symmetric model with 4 and 8 scatterers in the unitary regime
at various regions of the non-integrable system eigenstate labels
$\alpha$.\label{fig:NNDSpRigQ}}
\end{figure}

Thus, for all kinds of the model the statistics tend to the Wigner-Dyson
predictions on increase of the number of scatterers. This agrees with
the behavior of spectral rigidity of flat 2D billiards \cite{cheon1996}.
However, the present model does not demonstrate another property of
the 2D flat billiards --- the shifting toward Poisson statistics
at higher energy \cite{cheon1996}. It is clearly shown in Fig. \ref{fig:NNDSpRigQ},
where the plots for different energy regions are close together and
do not demonstrate a systematic dependence on the energy. This difference
is related to the nature of the logarithmic asymptotic freedom revealed
in\cite{cheon1996}. This effect is caused by the decreased effective
interaction strength $v_{\mathrm{eff}}\sim1/\ln\varepsilon$ (see
Eq. (19) in \cite{cheon1996}), while the characteristic energy level
separation $\partial\varepsilon_{\alpha}/\partial\alpha$ is independent
of the energy for 2D billiards with $\varepsilon_{\alpha}\propto\alpha$.
In contrast, if $\varepsilon_{\alpha}\propto\alpha^{\gamma}$ $(\gamma\neq1)$,
the derivation \cite{cheon1996} would lead to $v_{\mathrm{eff}}\propto\varepsilon^{1-1/\gamma}$,
while $\partial\varepsilon_{\alpha}/\partial\alpha\propto\varepsilon^{1-1/\gamma}$
has the same energy dependence and ratio of the effective interaction
strength to energy level separation is independent of energy. Therefore,
the logarithmic asymptotic freedom does not appear in the present
model with $\varepsilon_{\alpha}\propto\alpha^{2/3}$ as well as in
generic systems with $\varepsilon_{\alpha}\propto\alpha^{\gamma}$
$(\gamma\neq1)$, being a specific property of 2D billiards.
\begin{figure}[H]
\includegraphics[width=8cm]{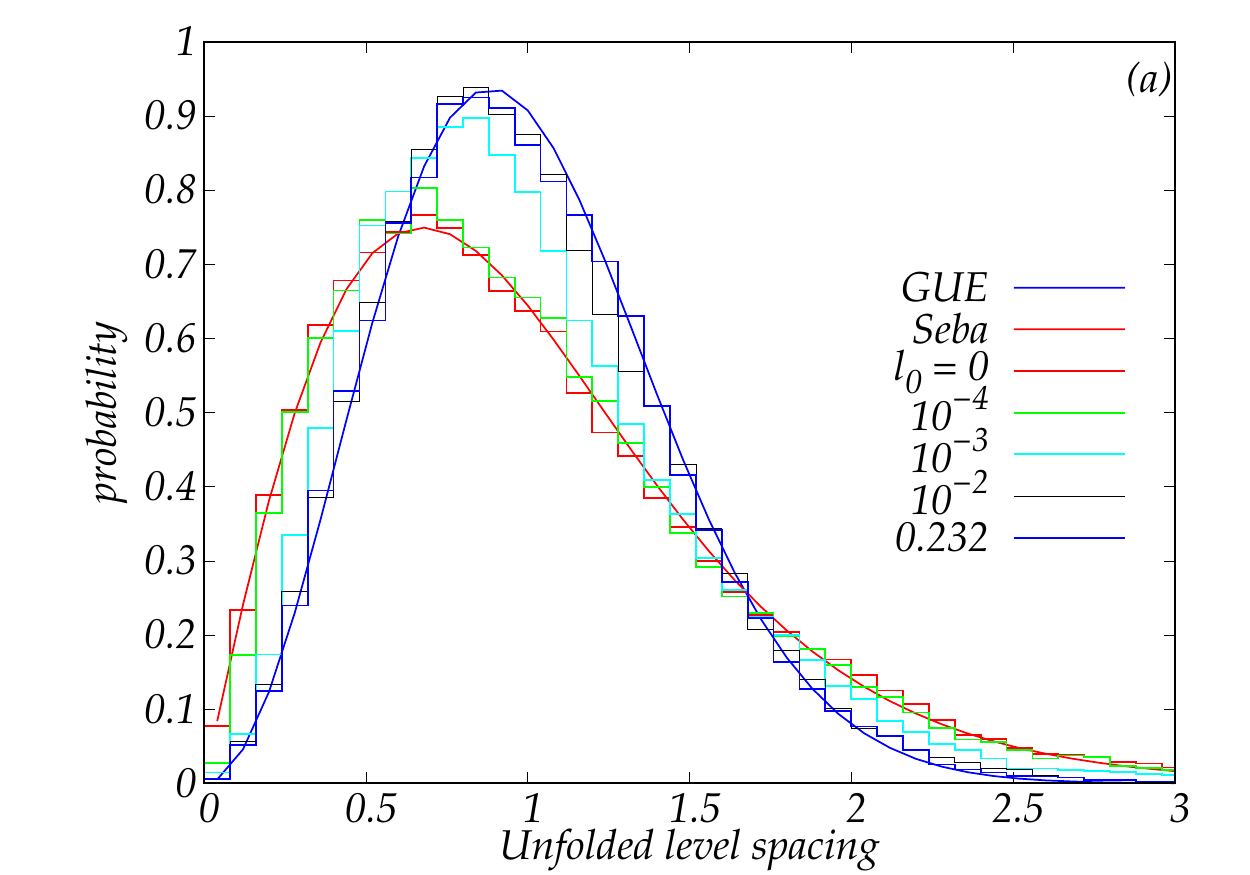}\includegraphics[width=8cm]{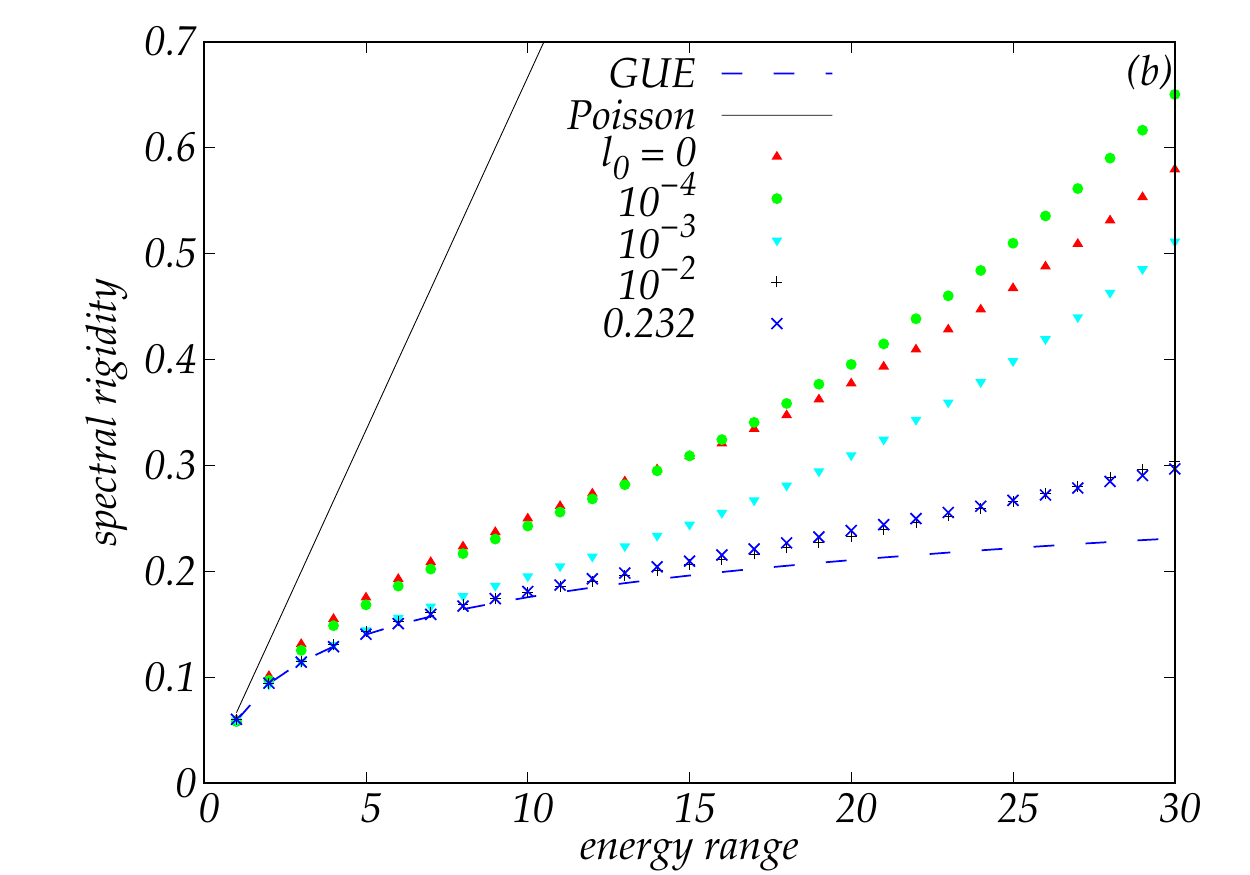}

\caption{Near-neighbor distribution (a) and the spectral rigidity (b) for the
non-symmetric model with 32 scatterers for various values of the scaled
vector potential $l_{0}$.\label{fig:NNDSpRigVP}}

\end{figure}

The transition between the T-invariant and non-symmetric models due
to the change of the vector potential is demonstrated in Fig. \ref{fig:NNDSpRigVP}.
The GUE and \v{S}eba statistics take place at $l_{0}<10^{-4}$ and
$l_{0}>10^{-2}$, respectively. 
\begin{figure}[H]
\includegraphics[width=8cm]{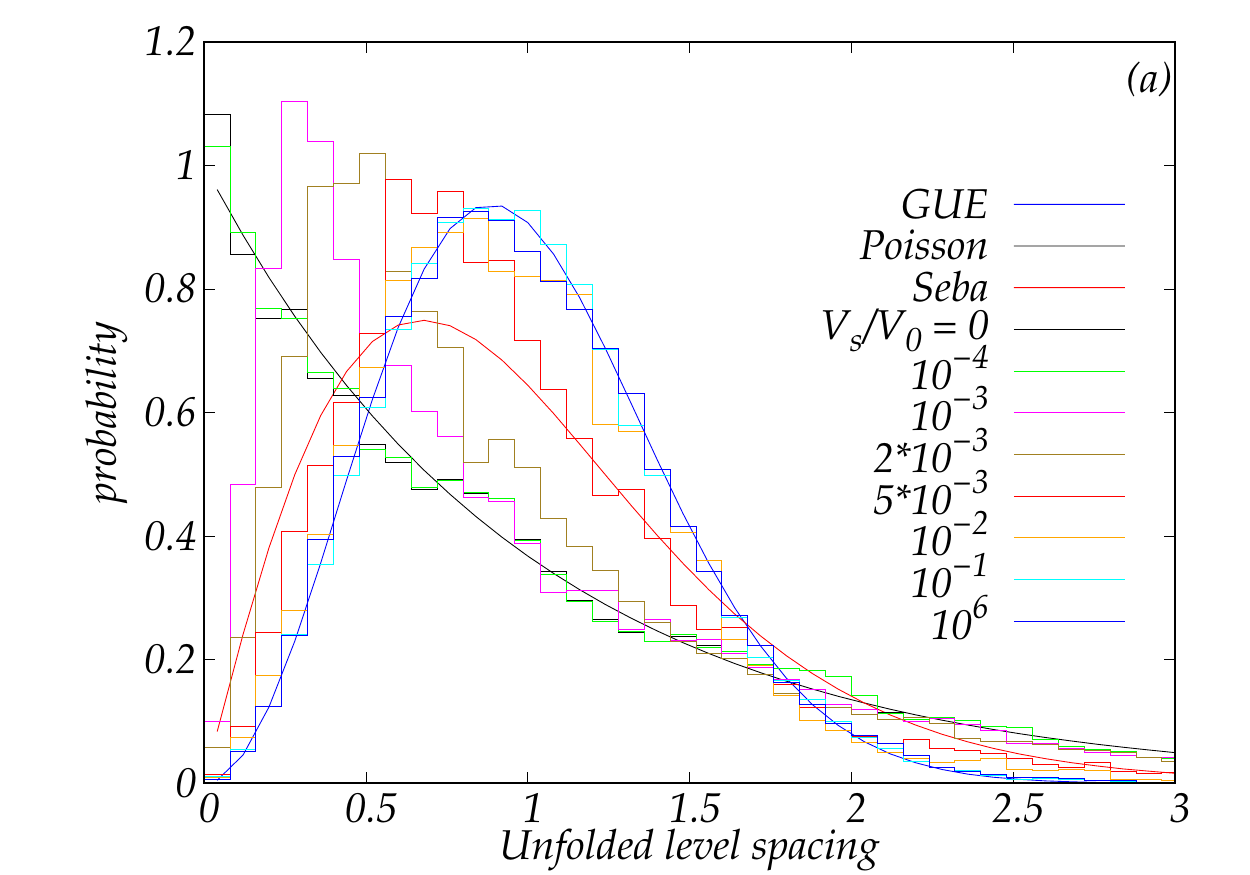}\includegraphics[width=8cm]{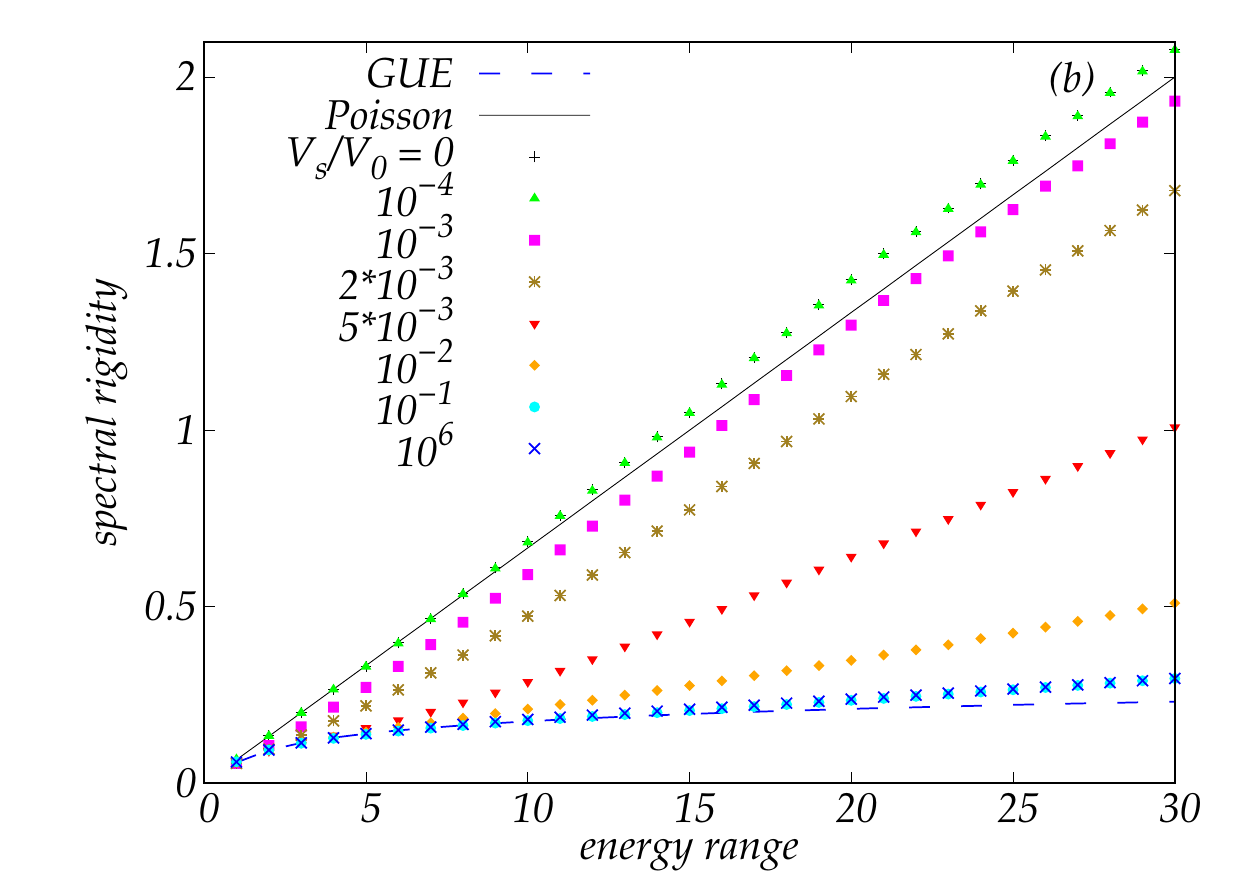}

\caption{Near-neighbor distribution (a) and the spectral rigidity (b) for the
non-symmetric model with 32 scatterers for various scatterer strengths.
\label{fig:NNDSpRigV}}
\end{figure}

The system chaoticity depends also on the scatterer strength $V_{s'}.$
NND approaches this unitary regime already at $V_{s'}=10^{-1}V_{0}$,
as Fig. \ref{fig:NNDSpRigV}(a) shows. For $V_{s'}=10^{-4}V_{0}$
NND almost coincides with the integrable system one. Spectral rigidity
demonstrates the same behavior (see Fig. \ref{fig:NNDSpRigV}(b)).
\begin{figure}[H]
\includegraphics[width=8cm]{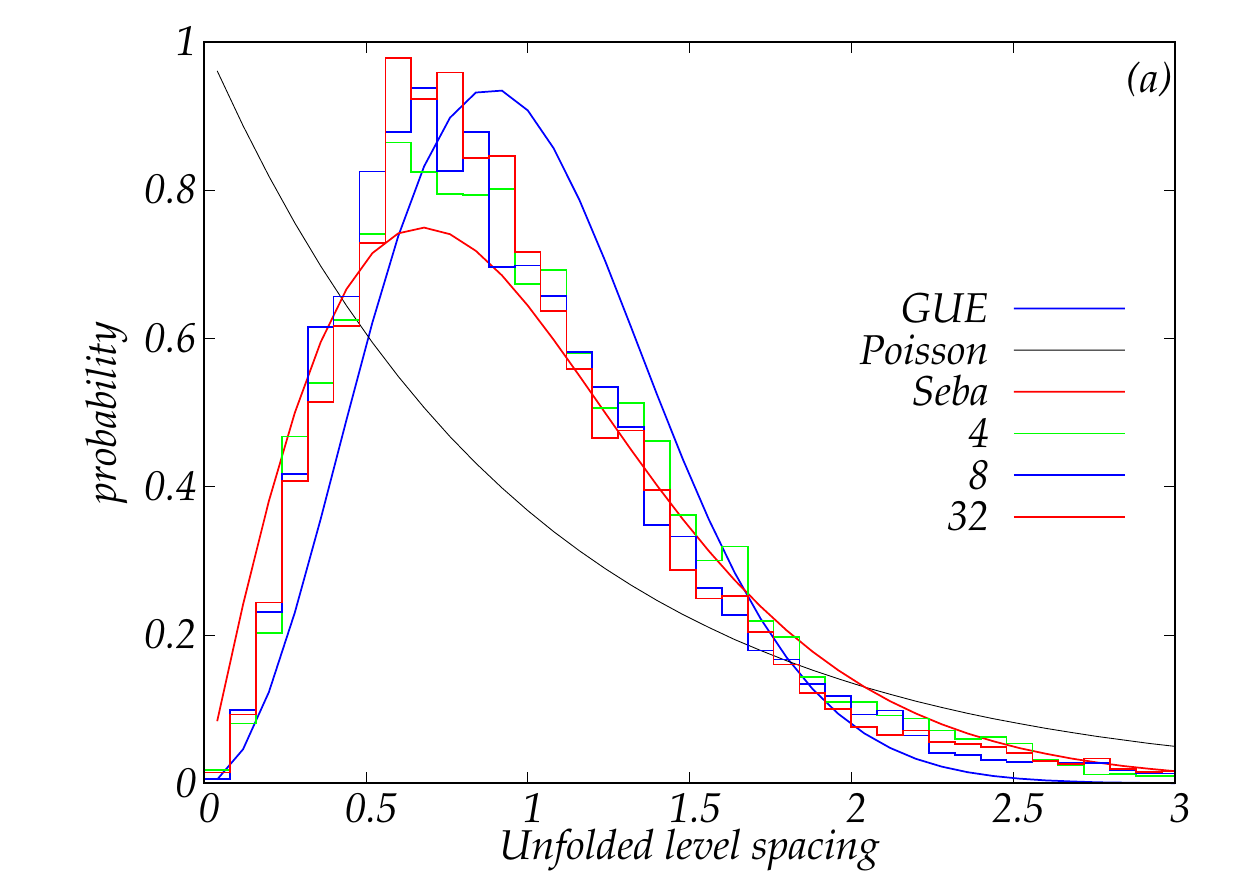}\includegraphics[width=8cm]{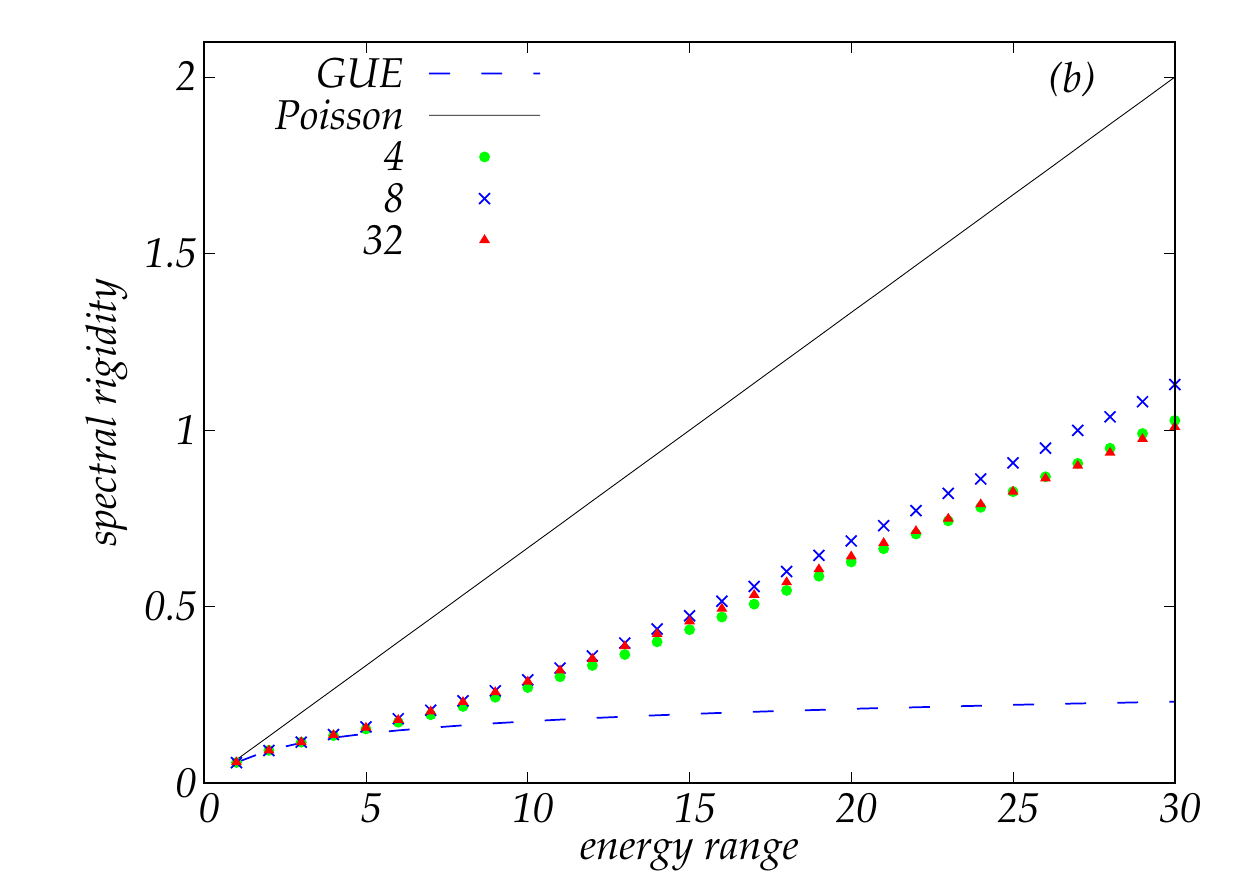}

\caption{Near-neighbor distribution (a) and the spectral rigidity (b) for the
non-symmetric model with 4, 8, and 32 scatterers at $V_{s'}/V_{0}=10^{-1}$,
$10^{-2}$, and $5\times10^{-3}$, respectively.\label{fig:NNDSpRigsv}}

\end{figure}

Thus, the system's chaotic properties depend on two parameters: the
number of scatterers and their strengths. Interaction of these parameters
is illustrated by Fig. \ref{fig:NNDSpRigsv}, which demonstrates that
the NND and spectral rigidity dependencies in the unitary regime for
4 scatterers are approached at $V_{s'}=10^{-2}V_{0}$ and $V_{s'}=5\times10^{-3}V_{0}$
for 8 and 32 scatterers, respectively.
\begin{figure}[H]
\includegraphics[width=8cm]{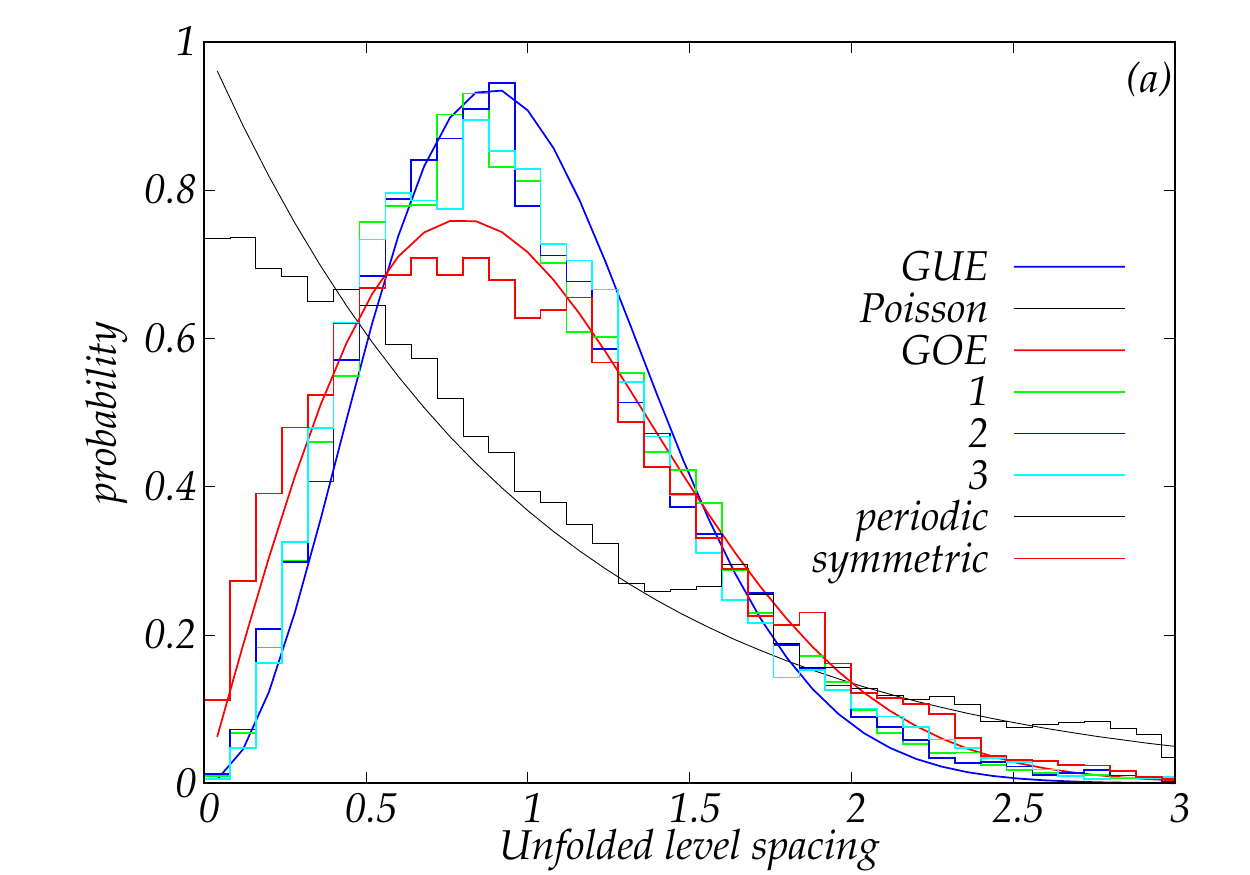}\includegraphics[width=8cm]{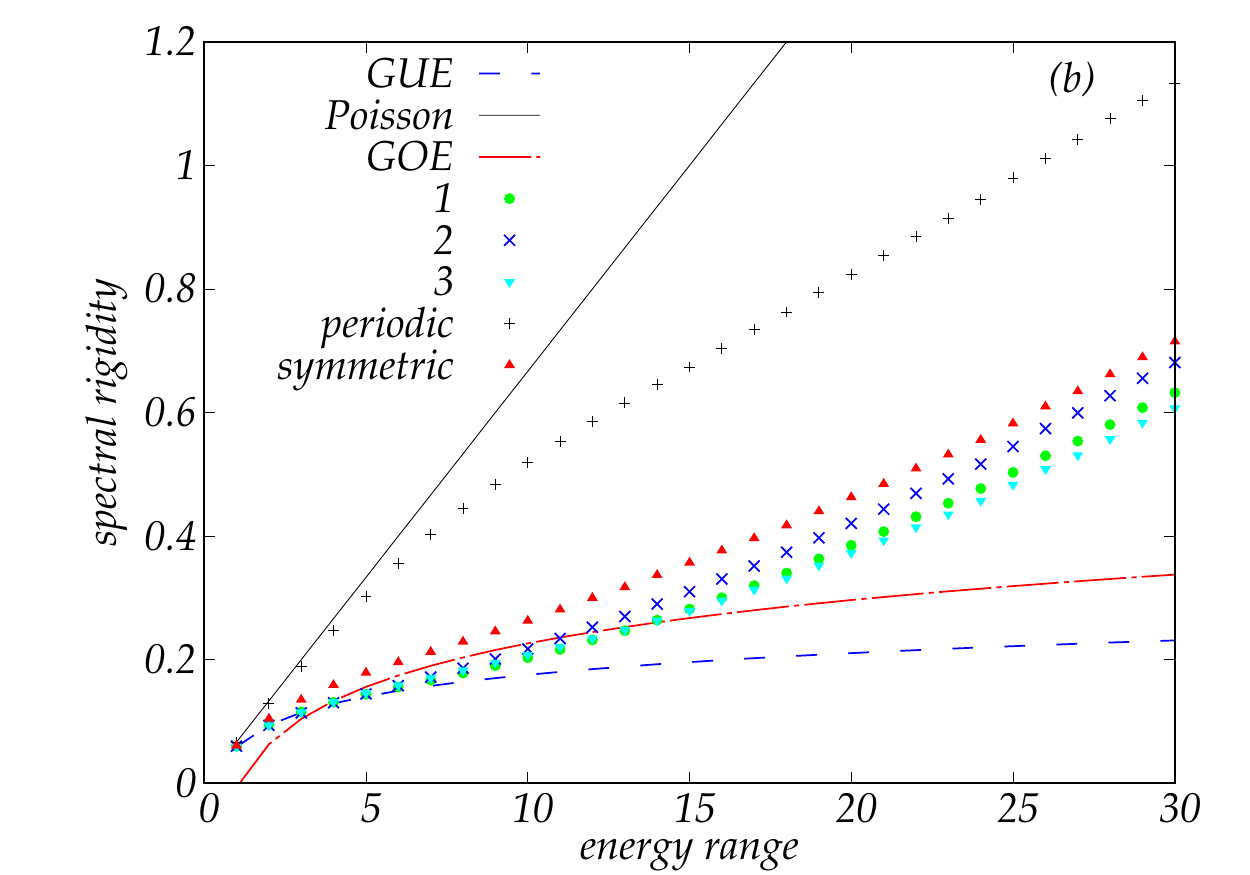}

\caption{Near-neighbor distribution (a) and the spectral rigidity (b) for the
non-symmetric model with 8 scatterers for various scatterer locations.
\label{fig:NNDSpRigP}}

\end{figure}

Figure \ref{fig:NNDSpRigP} shows dependence of the system statistics
on the scatterer locations. All non-symmetric cases (1 and 2, corresponding
to different sets of the random shifts $\delta_{s'}$ in \eqref{eq:zsnosymm},
and 3, where $\zeta_{1}=0$ and $\zeta_{s'}$ with $s'>1$ are chosen
randomly from the interval $[0,1]$ and sorted) provide close results
approaching the GUE predictions. The plots for the symmetric distribution
are clearly different and approach GOE predictions (see the discussion
above). 

If the scatterer positions form a periodic sequence, $\zeta_{s'}=(s'-1)/s$
and $V_{s'}$ is constant, the picture is completely different. In
this case, according to the Bloch's theorem, the eigenstate can be
expressed as $\left\langle \rho,z|\alpha\right\rangle =\left\langle \rho,z|\alpha_{p}\right\rangle \exp(ipz)$.
The $L$-periodicity plays the role of the Born-von Karman boundary
conditions, leading to the discrete spectrum of the quasimomentum
$p=2\pi k_{p}/L$ with integer $k_{p}$. The function $\left\langle \rho,z|\alpha_{p}\right\rangle $
has the period $L/s$ and satisfies the Schr\"{o}dinger equation
with single scatterer
\begin{equation}
\left(\hat{H}_{0}(A-p)+\hat{V}_{1}\right)\left|\alpha_{p}\right\rangle =E_{\alpha_{p}}\left|\alpha_{p}\right\rangle .
\end{equation}
Here the integrable Hamiltonian $\hat{H}_{0}(A-p)$ of the form \eqref{eq:H0}
contains the vector potential $A-p$. Therefore, the total energy
spectrum is a superposition of $s$ spectra of the one-scatterer systems
with scaled vector potentials $l_{0}-k_{p}$ ($k_{p}+s$ gives the
same result as $k_{p}$). This is the reason (see \cite{guhr1998})
why NND for the periodic case does not have a dip at small spacings
and both NND and spectral rigidity are close to the Poisson predictions.
\begin{figure}[H]
\includegraphics[width=8cm]{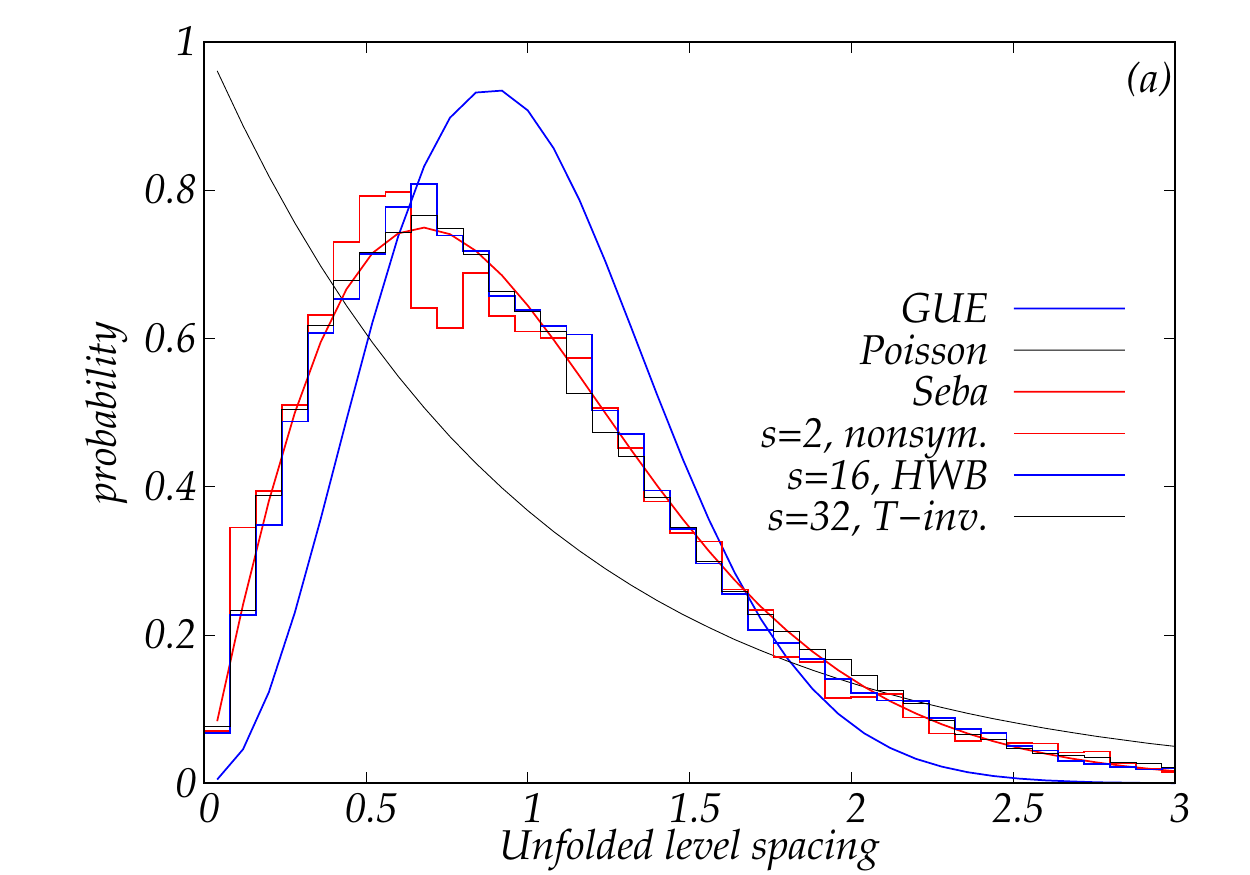}\includegraphics[width=8cm]{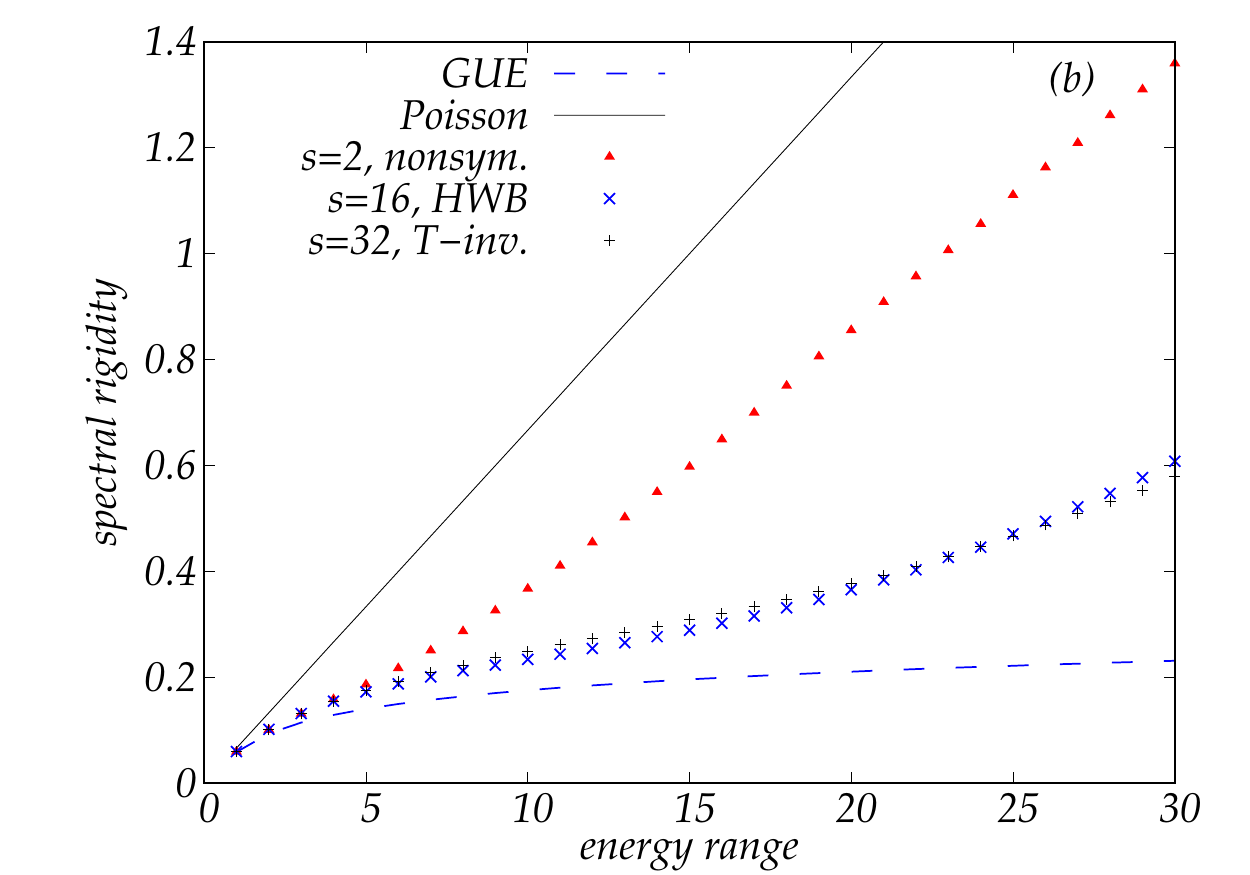}

\caption{Near-neighbor distributions (a) which are close to the \v{S}eba one
for various models and the corresponding spectral rigidity (b). \label{fig:NNDSpRigSeba}}

\end{figure}

The random matrix theory \cite{guhr1998,mehta,kota} predicts universal
spectral rigidity plots corresponding to the Poisson, GOE, and GUE
NNDs. However, the \v{S}eba NND can correspond to various spectral
rigidity plots, as it is shown in Fig. \ref{fig:NNDSpRigSeba}. It
is worth noting that the plots for the T-invariant PBC and HWB models
are close together, while the one for the T-noninvariant model is
completely different.

Statistics of energy spectra can be also characterized by average
level spacing ratio \cite{oganesyan2007,atas2013}which increases
with the system chaoticity. In the present case (see Appendix \ref{sec:Level-spacing-ratio})
, this monotonic increase takes place only in the vicinity of the
Poisson statistics. However, the average level spacing ratio becomes
almost the same for the \v{S}eba and GOE statistics and has strong
fluctuations on the transition between them. This may be related to
the small number of the degrees of freedom in the present models compared
to many-body models, where the level spacing ratio is generally used.
An additional advantage of the level spacing ratio is that unfolding
the spectrum is not required. However, this advantage is not essential
for the present models as the unfolding functions are well defined.
For these reasons, the level spacing ratio is not used here.

\section{Properties of wavefunctions \label{sec:PropWaveFun}}

Possibility of statistical description of quantum-chaotic systems
is based, through ETH, on properties of their wavefunctions. The number
of integrable system eigenstates comprising the non-integrable one
is characterized by the number of principal components (NPC) $\eta^{-1}$,
where $\eta=\sum_{nl}\left|\left\langle n,l|\alpha\right\rangle \right|^{4}$
is IPR. Equations \eqref{eq:LippShw} and \eqref{eq:nlVsalpha} allow
us to express the expansion coefficients here in the form
\begin{equation}
\left\langle n,l|\alpha\right\rangle =\sqrt{\mathcal{N}_{\alpha}}\frac{1}{\varepsilon_{\alpha}-\varepsilon_{nl}}\sum_{s'=1}^{s}\frac{V_{s'}}{V_{0}}e^{-2i\pi l\zeta_{s'}}\left\langle \mathbf{R}_{s'}|\alpha\right\rangle _{reg},\label{eq:nlalpha}
\end{equation}
where $\left\langle \mathbf{R}_{s'}|\alpha\right\rangle _{reg}$ are
solutions to the system \eqref{eq:linsysreg} and the normalization
factor $\mathcal{N}_{\alpha}$ is determined by the normalization
condition $\sum_{nl}\left|\left\langle n,l|\alpha\right\rangle \right|^{2}=1$.
For the energies \eqref{eq:Enl} the sums over $n$ here and in IPR
can be expressed in terms of the Hurwitz zeta functions (see \cite{DLMF})
\begin{equation}
\sum_{n=0}^{\infty}\frac{1}{(\varepsilon_{\alpha}-\varepsilon_{nl})^{k}}=\frac{(-1)^{k}}{\lambda^{k}}\zeta(k,q_{l}),\label{eq:Sumnzeta}
\end{equation}
where 
\begin{equation}
q_{l}=\frac{\pi^{2}(l-l_{0})^{2}-\varepsilon_{\alpha}}{\lambda}.\label{eq:ql}
\end{equation}
Then the normalization condition takes the form $\sum_{l=-\infty}^{\infty}P_{l}=1$,
where
\begin{equation}
P_{l}\equiv\sum_{n=0}^{\infty}\left|\left\langle n,l|\alpha\right\rangle \right|^{2}=\frac{\mathcal{N}_{\alpha}}{\lambda^{2}}\Lambda_{l}\zeta(2,q_{l})\label{eq:Pl}
\end{equation}
is the occupation of the states with the given axial quantum number
$l$ and
\begin{equation}
\Lambda_{l}=\left|\sum_{s'=1}^{s}\frac{V_{s'}}{V_{0}}e^{-2i\pi l\zeta_{s'}}\left\langle \mathbf{R}_{s'}|\alpha\right\rangle _{reg}\right|^{2}.
\end{equation}
Similarly, for IPR we have
\begin{equation}
\eta\equiv\sum_{l=-\infty}^{\infty}\sum_{n=0}^{\infty}\left|\left\langle n,l|\alpha\right\rangle \right|^{4}=\frac{\mathcal{N}_{\alpha}^{2}}{\lambda^{4}}\sum_{l=-\infty}^{\infty}\Lambda_{l}^{2}\zeta(4,q_{l}).
\end{equation}
The expressions above are used for T-noninvariant models (non-symmetric
and symmetric), where $A\neq0$ and $\left\langle \mathbf{R}_{s'}|\alpha\right\rangle _{reg}$
are complex. In the T-invariant models ($A=0$) $\left\langle \mathbf{R}_{s'}|\alpha\right\rangle _{reg}$
are real. For PBC the normalization condition can be expressed as
$\sum_{l=0}^{\infty}P_{l}^{T}=1$, where
\begin{equation}
P_{l}^{T}=\frac{\mathcal{N}_{\alpha}}{\lambda^{2}}(2-\delta_{l0})(\Lambda_{l}^{c}+\Lambda_{l}^{s})\zeta(2,q_{l})\label{eq:PlT}
\end{equation}
and
\begin{equation}
\Lambda_{l}^{c,s}=\left(\sum_{s'=1}^{s}\frac{V_{s'}}{V_{0}}\left\langle \mathbf{R}_{s'}|\alpha\right\rangle _{reg}\left\{ \begin{array}{c}
\cos2\pi l\zeta_{s'}\\
\sin2\pi l\zeta_{s'}
\end{array}\right\} \right)^{2}.
\end{equation}
Respectively, IPR can be expressed as
\begin{equation}
\eta=\frac{\mathcal{N}_{\alpha}^{2}}{\lambda^{4}}\sum_{l=0}^{\infty}(2-\delta_{l0})(\Lambda_{l}^{c}+\Lambda_{l}^{s})^{2}\zeta(4,q_{l}).
\end{equation}

For HWB we have the normalization condition $\sum_{l=1}^{\infty}P_{l}^{B}=1$
with
\begin{equation}
P_{l}^{B}=\frac{\mathcal{N}_{\alpha}}{\lambda^{2}}\Lambda_{l}^{B}\zeta(2,q_{l}^{B})\label{eq:PlB}
\end{equation}
and
\begin{equation}
\eta=\frac{\mathcal{N}_{\alpha}^{2}}{\lambda^{4}}\sum_{l=1}^{\infty}\left(\varLambda_{l}^{B}\right)^{2}\zeta(4,q_{l}^{B}),
\end{equation}
where
\begin{equation}
\Lambda_{l}^{B}=\left(\sum_{s'=1}^{s}\frac{V_{s'}}{V_{0}}\left\langle \mathbf{R}_{s'}|\alpha\right\rangle _{reg}\sin\pi l\zeta_{s'}\right)^{2}
\end{equation}
and $q_{l}^{B}=(\pi^{2}l^{2}/4-\varepsilon_{\alpha})/\lambda$.
\begin{figure}[H]
\includegraphics[width=8cm]{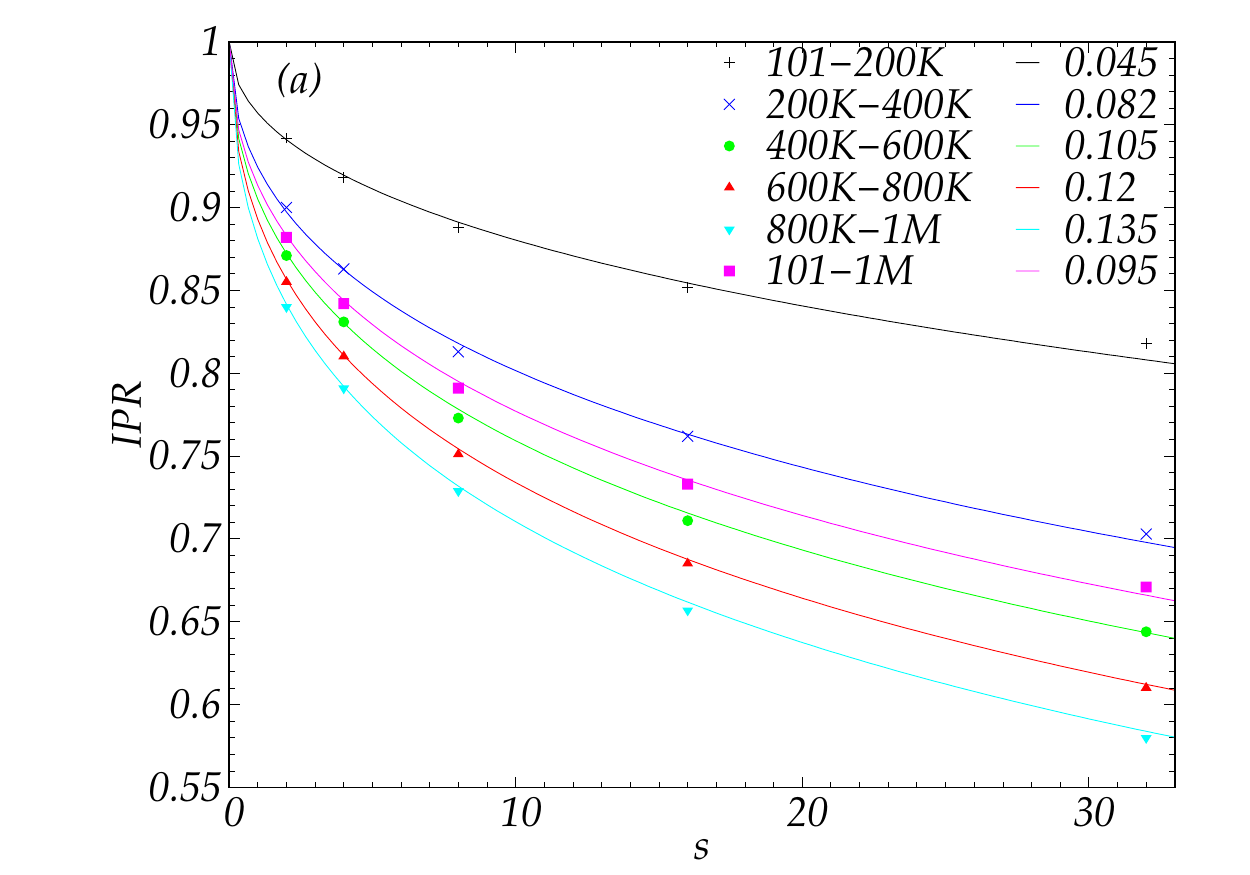}\includegraphics[width=8cm]{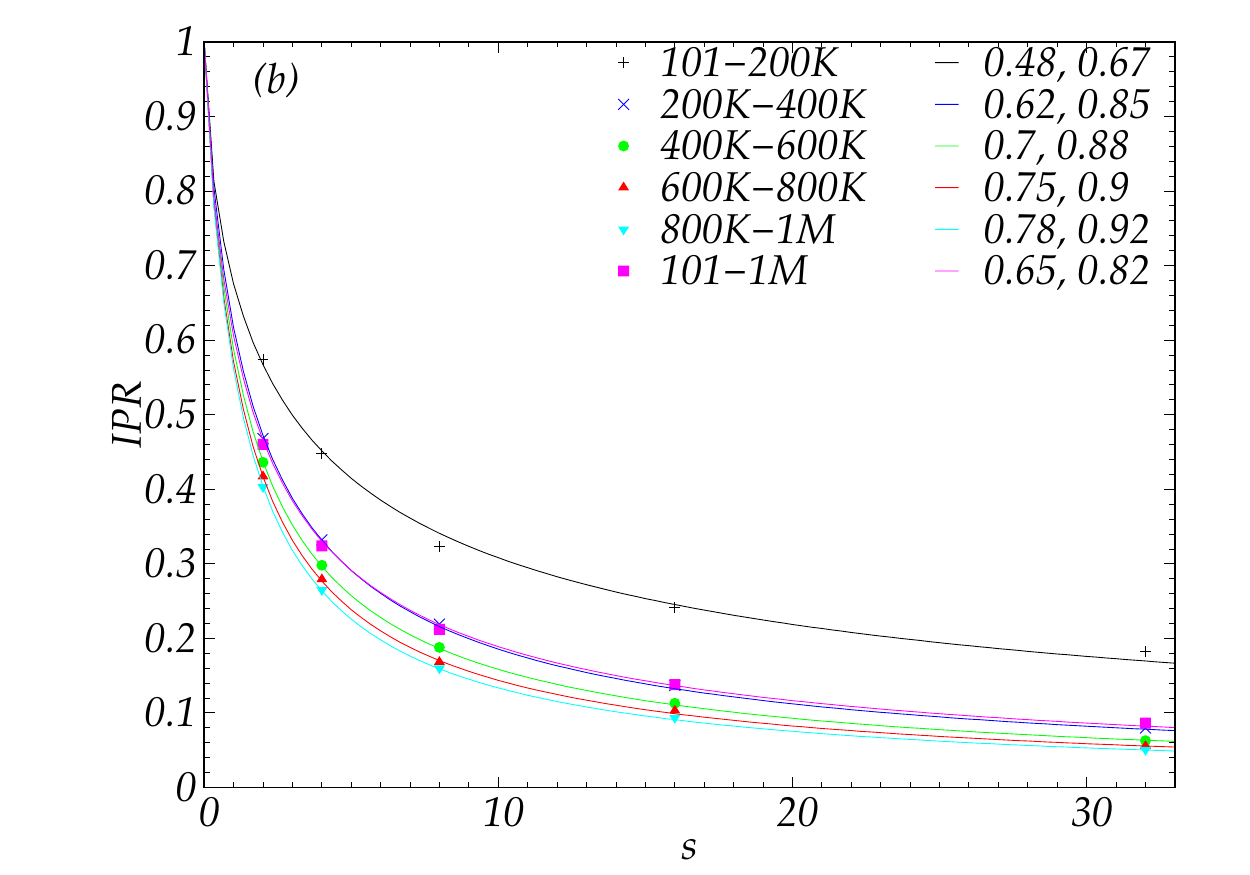}

\caption{Inverse participation ratio as a function of the number of scatterers
calculated for the non-symmetric model with $V_{s'}/V_{0}=10^{-3}$
(a) and $V_{s'}/V_{0}=10^{-2}$ (b) at various regions of the non-integrable
system eigenstate labels $\alpha$. The lines show the dependencies
$\eta_{s}=1/(1+\nu's^{0.48})$ with the $\nu'$ values presented in
the legend in the part (a) and $\eta_{s}=1/(1+\nu's^{\gamma})$ with
the $\nu'$ and $\gamma$ values presented in the legend in the part
(b).\label{fig:IPRnscat}}
\end{figure}

\begin{figure}[H]
\includegraphics[width=8cm]{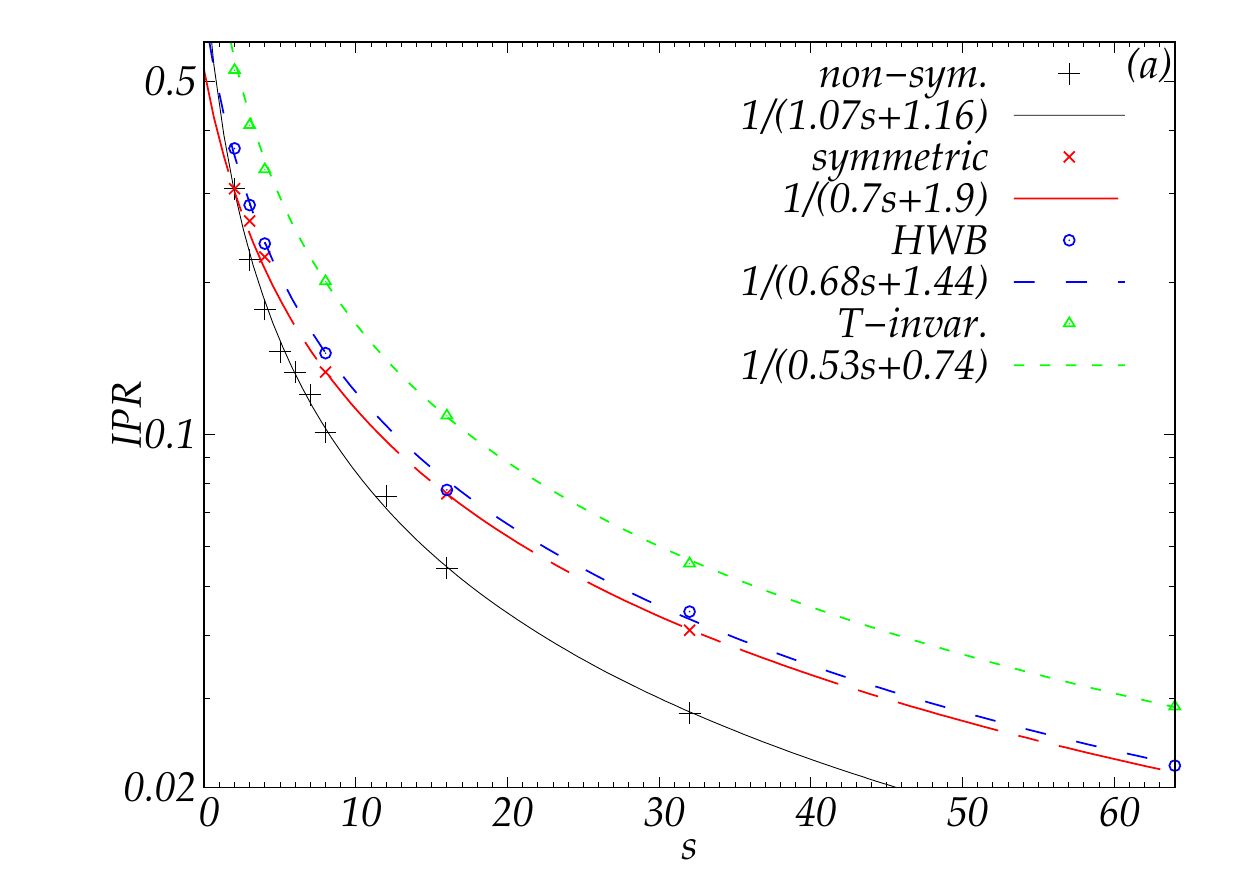}\includegraphics[width=8cm]{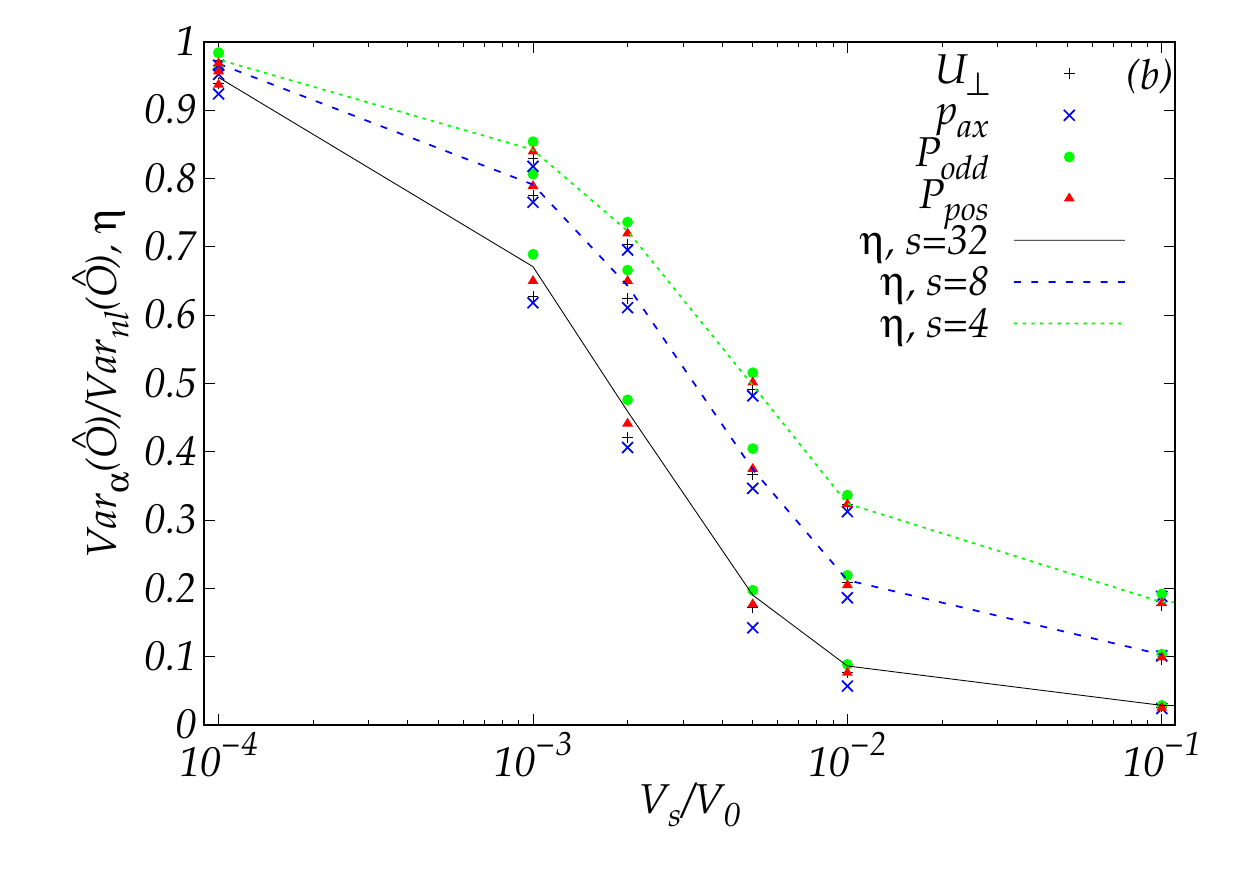}

\caption{(a) Inverse participation ratio as a function of the number of scatterers
calculated in the unitary regime for four kinds of the model. The
lines show the best fit by inversely linear functions. This part uses
the same data as Fig. 3(a) in \cite{yurovsky2023}. (b) Ratio of fluctuation
variances between the non-integrable and integrable systems eigenstates
as a function of the scatterer strength for the non-symmetric model
with 4, 8, and 32 scatterers. The points correspond to the four observables,
the lines connect the calculated IPR values. The data for 32 scatterers
are the same as in Fig. 3(b) of \cite{yurovsky2023}. \label{fig:FlicV}}
\end{figure}

A recurrence relation $\eta_{s}^{-1}=\eta_{s-1}^{-1}+\nu$ was derived
\cite{yurovsky2023} for NPC $\eta_{s}^{-1}$ of the system with $s$
scatterers. This means that NPC increases and, respectively, IPR decreases
with the number of scatterers. In the case of weak interaction the
dependence of NPC on the number of scatterers is nonlinear (see Fig.
\ref{fig:IPRnscat}). This is a consequence of the strong dependence
of the system's chaotic properties on the number of scatterers. NPC
also increases with the eigenstate energy due to increase of the energy
level density. In the case of the statistics of energy spectra, this
increase was compensated by decrease of the effective interaction
strength (see Fig. \ref{fig:NNDSpRigQ} and the related discussion
above). Here we see that the wavefunction properties are determined
by the interaction strength $V_{s'}$ rather than the effective one.
The NPC dependence on the number of scatterers can be approximated
by $\eta_{s}^{-1}=1+\nu's^{\gamma}$. For a weak interaction $V_{s'}=10^{-3}V_{0}$
the power $\gamma\approx0.48$ becomes independent of the eigenstate
energy (see Fig. \ref{fig:IPRnscat}(a)). For stronger interaction
$V_{s'}/V_{0}=10^{-2}$ (see Fig. \ref{fig:IPRnscat}(b)) the power
$\gamma$ increases with the eigenstate energy and the dependence
of NPC on $s$ tends to the linear one. This means that $\nu$ is
independent of $s$ since the system's chaotic properties are independent
of the number of scatterers. In the unitary regime, this dependence
is confirmed by IPR calculated for all kinds of the model (see Fig.
\ref{fig:FlicV}(a)) which is approximated by inverse-linear functions
with a good accuracy. We can see that for each number of scatterers
the non-symmetric model has the minimal IPR and, therefore, demonstrates
the highest chaoticity, the T-invariant PBC model has the highest
IPR, and the HWB one lies between them. This order agrees with the
NND and spectral rigidity of energy spectra for these models discussed
in Sec. \ref{sec:Energy-spectra} above. However, the symmetric T-noninvariant
model has substantially higher IPR than the non-symmetric one, although
properties of energy spectra of these models demonstrate similar chaoticity.
This difference can be related to properties of real and complex random
Gaussian variables \cite{yurovsky2023}. As well as any characteristic
of chaos, IPR depends also on the interaction strength (see Fig. \ref{fig:FlicV}(b)).
This figure also demonstrates that the systems with 4, 8, and 32 scatterers
have approximately the same IPR ($\eta\approx0.2$) at $V_{s'}/V_{0}=10^{-1}$(and
in the unitary regime), $10^{-2}$, and $5\times10^{-3},$respectively,
in agreement with the energy spectra statistics (see Fig. \ref{fig:NNDSpRigsv}).

Chaotic properties of physical systems are also characterized by fluctuations
of observable expectation values. Expectation value of the observable
$\hat{O}$ in eigenstates of the non-integrable system is related
to ones in integrable system eigenstates 
\begin{equation}
\left\langle \alpha\left|\hat{O}\right|\alpha\right\rangle =\sum_{n,l,n',l'}\left\langle \alpha|n',l'\right\rangle \left\langle n',l'\left|\hat{O}\right|n,l\right\rangle \left\langle n,l|\alpha\right\rangle ,\label{eq:ExpValalphan}
\end{equation}
where the expansion coefficients $\left\langle n,l|\alpha\right\rangle $
are given by \eqref{eq:nlalpha}.

Four observables are considered here. The transverse potential energy
$m\omega_{\perp}^{2}\rho^{2}/2$ is non-diagonal in the integrable
system eigenstates
\begin{equation}
\left\langle n',l'\left|\frac{1}{2}m\omega_{\perp}^{2}\rho^{2}\right|n,l\right\rangle =\frac{\omega_{\perp}}{2}\delta_{ll'}\left[\left(2n+1\right)\delta_{nn'}-n\delta_{n'n-1}-n'\delta_{nn'-1}\right].\label{eq:ExpValPotn}
\end{equation}
As the potential energy increases with the total energy, the part
$U_{\perp}$ of the transverse potential energy in the total energy
is considered here. Its expectation value in the non-integrable system
eigenstates can be expressed as (see Appendix \ref{sec:Expectation-values})
\begin{equation}
\left\langle \alpha\left|U_{\perp}\right|\alpha\right\rangle =\frac{\omega_{\perp}}{2E_{\alpha}}\left[1+2\frac{\mathcal{N}_{\alpha}}{\lambda^{2}}\sum_{l=-\infty}^{\infty}\Lambda_{l}\left(1-q_{l}\zeta(2,q_{l})\right)\right]\label{eq:ExpValUperp}
\end{equation}
for T-noninvariant models. In the T-invariant PBC case we have
\begin{equation}
\left\langle \alpha\left|U_{\perp}\right|\alpha\right\rangle =\frac{\omega_{\perp}}{2E_{\alpha}}\left[1+2\frac{\mathcal{N}_{\alpha}}{\lambda^{2}}\sum_{l=0}^{\infty}(2-\delta_{l0})(\Lambda_{l}^{c}+\Lambda_{l}^{s})\left(1-q_{l}\zeta(2,q_{l})\right)\right].\label{eq:ExpValUperpT}
\end{equation}
In the last case, HWB, the expectation value takes the form
\begin{equation}
\left\langle \alpha\left|U_{\perp}\right|\alpha\right\rangle =\frac{\omega_{\perp}}{2E_{\alpha}}\left[1+2\frac{\mathcal{N}_{\alpha}}{\lambda^{2}}\sum_{l=1}^{\infty}\Lambda_{l}^{B}\left(1-q_{l}^{B}\zeta(2,q_{l}^{B})\right)\right].\label{eq:ExpValUperpB}
\end{equation}

Other observables, diagonal in integrable system eigenstates, are
the axial momentum $\left\langle nl\left|\hat{p}_{ax}\right|n'l'\right\rangle =l\delta_{n'n}\delta_{l'l}$,
the occupation of positive momenta $\left\langle nl\left|\hat{P}_{pos}\right|n'l'\right\rangle =\delta_{n'n}\delta_{l'l}\theta(l)$,
where $\theta(l)=0$ for $l<0$, $1/2$ for $l=0$, and $1$ for $l>0$,
and the occupation of the odd axial modes $\left\langle nl\left|\hat{P}_{odd}\right|n'l'\right\rangle =\delta_{n'n}\delta_{l'l}\delta_{l\mathrm{mod}2,1}$,
where $l\mathrm{mod}2$ is the reminder of the division of $l$ by
2. For T-noninvariant models their expectation values are expressed
in terms of the occupations $P_{l}$ \eqref{eq:Pl}, 
\begin{equation}
\left\langle \alpha\left|\hat{p}_{ax}\right|\alpha\right\rangle =\sum_{l=-\infty}^{\infty}lP_{l},\quad\left\langle \alpha\left|\hat{P}_{pos}\right|\alpha\right\rangle =\frac{1}{2}P_{0}+\sum_{l=1}^{\infty}P_{l},\quad\left\langle \alpha\left|\hat{P}_{odd}\right|\alpha\right\rangle =\sum_{l=-\infty}^{\infty}P_{2l+1}.
\end{equation}
For T-invariant models, $\left\langle \alpha\left|\hat{p}_{ax}\right|\alpha\right\rangle =0$
and $\left\langle \alpha\left|\hat{P}_{pos}\right|\alpha\right\rangle =1/2$
do not fluctuate, while $\left\langle \alpha\left|\hat{P}_{odd}\right|\alpha\right\rangle =\sum_{l=1}^{\infty}P_{2l-1}^{T,B}$
are expressed in terms of probabilities \eqref{eq:PlT} and \eqref{eq:PlB},
respectively.

\begin{figure}[H]
\includegraphics[width=8cm]{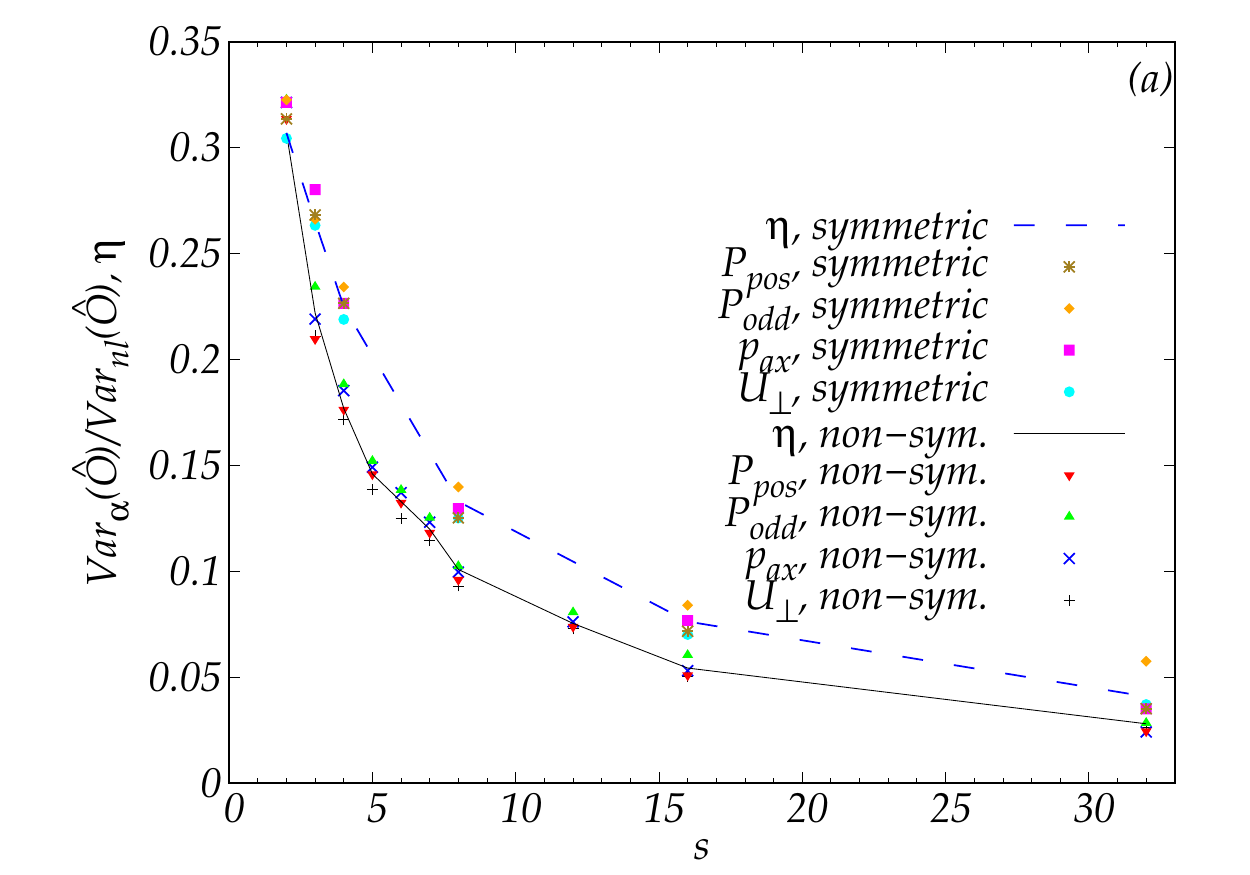}\includegraphics[width=8cm]{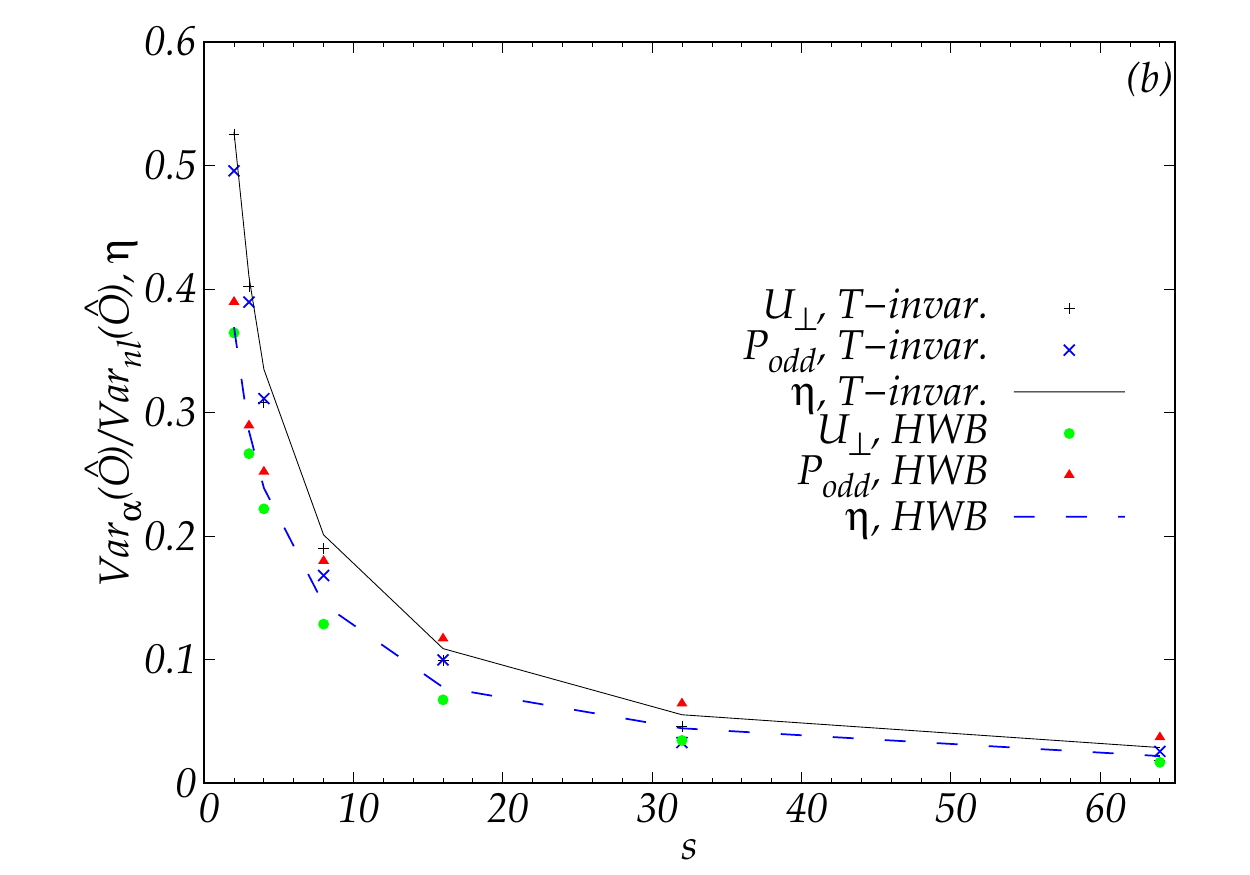}

\caption{Ratio of fluctuation variances between the non-integrable and integrable
systems eigenstates as a function of the number of scatterers for
T-noninvariant (a) and T-invariant (b) models in the unitary regime.
The points correspond to the four observables, the lines connect the
calculated IPR values. The data for the non-symmetric model in the
part (a) are the same as in Fig. 3(c) of \cite{yurovsky2023}. \label{fig:Flucnscatu}}
\end{figure}

For an observable $\hat{O}$, the variance of its expectation value
fluctuations between non-integrable system eigenstates is defined
as
\begin{equation}
\mathrm{Var}_{\alpha}(\hat{O})=\overline{\left\langle \alpha\left|\hat{O}\right|\alpha\right\rangle ^{2}}-\overline{\left\langle \alpha\left|\hat{O}\right|\alpha\right\rangle }^{2}.
\end{equation}
According to \cite{neuenhahn2012}, this variance is proportional
to IPR and the variance between the integrable system eigenstates
\begin{equation}
\mathrm{Var}_{\alpha}(\hat{O})=\eta\mathrm{Var}_{nl}(\hat{O}).\label{eq:VarOIPR}
\end{equation}
The latter variances are calculated in Appendix for the four observables
presented above. The variance of the axial momentum 
\begin{equation}
\mathrm{Var}_{nl}(p_{ax})=(\varepsilon_{max}^{5/2}-\varepsilon_{min}^{5/2})/[5\pi^{2}(\varepsilon_{max}^{3/2}-\varepsilon_{min}^{3/2})]\label{eq:Varnlpax}
\end{equation}
 depends on the averaging interval $[\varepsilon_{min},\varepsilon_{max}]$
boundaries. The variances of other observables are independent of
the interval, $\mathrm{Var}_{nl}(\hat{P}_{pos})=1/4$, $\mathrm{Var}_{nl}(\hat{P}_{odd})=1/4$,
and $\mathrm{Var}_{nl}(\hat{U}_{\perp})=1/45$. Figure \ref{fig:FlicV}(b)
confirms the rule \eqref{eq:VarOIPR} for the integrability-chaos
transition on variation of the scatterer strength in the non-symmetric
model, both for 4, 8, and 32 scatterers. This rule is also confirmed
when the number of scatterers is changed for all four models considered
here (see Fig. \eqref{fig:Flucnscatu}).

\section{Conclusion }

An effective method of numerical solution, based on properties of
high-rank separable perturbations, is developed for a harmonic waveguide
with a vector potential and either PBC or HWB in the axial direction,
perturbed by zero-range scatterers along the waveguide axis. The energy-degeneracy
of the unperturbed system can be lifted by the vector potential which
also lifts T-invariance. The energy spectra properties --- near-neighbor
distribution and spectral rigidity, as well as IPR and fluctuation
variance of observable expectation values, are calculated for $10^{6}$
eigenstates. The chaoticity measures of the model increase with the
number of scatterers and their strengths. This allows exploring the
integrability-chaos transition. 

In T-noninvariant models, the energy spectra properties follow the
\v{S}eba plots already for 2 scatterers and approach the Wigner-Dyson
predictions for 32 scatterers. The model with non-symmetric scatterer
distribution approaches the GUE statistics, while the P-invariant
distribution leads to the GOE statistics inherent in T-invariant systems.
It is a consequence of PT-invariance of the latter model, leading
to a real interaction matrix. Similarly, the IPR difference between
the two kinds of models can be related to properties of real and complex
wavefunctions. 

The T-invariant HWB and PBC models approach the \v{S}eba statistics
only for 16 and 32 scatterers, respectively, and the GOE one for 32
and 64 scatterers, respectively, i.e., much slower than the T-noninvariant
models. This can be related to the vector potential, which randomizes
the sequence of quantum numbers of energy-ordered eigenstates in the
integrable system. 

Calculation for different numbers of scatterers and their strengths
confirm the prediction \cite{yurovsky2023} that IPR decreases with
the number of scatterers. The dependence is inversely proportional
for strong scatterers. The prediction \cite{neuenhahn2012} that the
ratio of the observable fluctuation variances for the nonintegrable
and integrable systems is approximately equal to IPR is confirmed
as well. Thus, all criteria of chaoticity confirm that the model approaches
the complete quantum chaos and the eigenstate thermalization when
the number of scatterers is increased.


\appendix

\section{Derivation of the summands $T_{n}(\zeta_{s'},\zeta_{s''})$ and $T_{n}^{reg}$
in Eq. \eqref{eq:Sspspp}\label{sec:DerivationT}}

Let us define 
\begin{equation}
T_{n}(\frac{z}{L},\frac{z'}{L})=\frac{2\pi a_{\perp}^{2}}{mL}\sum_{l}\frac{\left\langle 0,0,z|nl\right\rangle \left\langle nl|0,0,z'\right\rangle }{E-E_{nl}}.\label{eq:Tndef}
\end{equation}
For the PBC models using Eqs. \eqref{eq:rhon}, \eqref{eq:zlpbc},
and \eqref{eq:Enl} we get
\begin{equation}
T_{n}(\zeta,\zeta')=\sum_{l=-\infty}^{\infty}\frac{\exp(2i\pi l(\zeta-\zeta'))}{\varepsilon-\lambda n-\pi^{2}(l-l_{0})^{2}},\label{eq:TnPBC}
\end{equation}
where $\varepsilon=mL^{2}(E-\omega_{\perp})/2$. Due to translational
invariance of PBC, $T_{n}$ is a function of $z-z'$ only. Then $T_{n}(\zeta_{s'},\zeta_{s''})=T_{n}(\zeta_{s'}-\zeta_{s''},0)$,
and, therefore, only $T_{n}(\zeta,0)$ should be evaluated. Farther,
the partial fraction decomposition 
\begin{equation}
\frac{1}{\varepsilon-\lambda n-\pi^{2}(l-l_{0})^{2}}=\frac{1}{2\pi p_{n}}\left(\frac{1}{l-l_{0}+p_{n}/\pi}-\frac{1}{l-l_{0}-p_{n}/\pi}\right),
\end{equation}
where $p_{n}=\sqrt{\varepsilon-\lambda n}$, allows us to use the
summation formula 
\begin{equation}
\sum_{l=-\infty}^{\infty}\frac{\exp(2i\pi l\zeta)}{l+a}=\frac{\pi}{\sin\pi a}\exp(-2i\pi a(\zeta-[\zeta]-1/2))\label{eq:summlexp}
\end{equation}
following from Eq. (5.4.3.4) in \cite{prudnikov}. As $0\leq\zeta<1$,
this leads to
\begin{equation}
T_{n}(\zeta,0)=\frac{1}{2p_{n}}e^{2i\pi l_{0}\zeta}\left[e^{2ip_{n}\zeta}\left(\cot(\pi l_{0}+p_{n})-i\right)-e^{-2ip_{n}\zeta}\left(\cot(\pi l_{0}-p_{n})-i\right)\right].\label{eq:Tn}
\end{equation}
In the diagonal elements of the matrix $S_{s's''}(\varepsilon)$ {[}see
Eq. \eqref{eq:Sspspp}{]} we need
\begin{equation}
T_{n}(0,0)=\frac{\sin2p_{n}}{p_{n}(\cos2\pi l_{0}-\cos2p_{n})}.\label{eq:Tn0}
\end{equation}
When $\lambda n>\varepsilon_{\alpha}$, $p_{n}$ becomes imaginary,
$|p_{n}|=\sqrt{\lambda n-\varepsilon_{\alpha}}$, and we have
\begin{equation}
T_{n}(\zeta,0)=-\frac{1}{|p_{n}|}e^{2i\pi l_{0}\zeta}\left(\frac{e^{-2|p_{n}|\zeta}}{1-e^{2i\pi l_{0}-2|p_{n}|}}+\frac{e^{-2|p_{n}|(1-\zeta)}}{e^{2i\pi l_{0}}-e^{-2|p_{n}|}}\right).\label{eq:Tnc}
\end{equation}
In the limit of the large $n$ and for any $0<\zeta<1$ the two terms
in the parentheses decay as $\exp(-2\sqrt{\lambda n}\zeta)$ and $\exp(-2\sqrt{\lambda n}(1-\zeta))$,
respectively. However, if $\zeta=0$, $T_{n}(0,0)\sim n^{-1/2}$ and
the sum of $T_{n}(0,0)$ diverges. In order to regularize this sum,
let us represent $T_{n}(\zeta,0)$ in the limit of $\zeta\rightarrow0$
as
\begin{equation}
T_{n}(\zeta,0)\sim-\frac{e^{-2|p_{n}|\zeta}}{|p_{n}|}+T_{n}^{reg},\quad T_{n}^{reg}=-\frac{2}{|p_{n}|}\frac{(\cos2\pi l_{0}-e^{-2|p_{n}|})e^{-2|p_{n}|}}{(e^{-2|p_{n}|}-2\cos2\pi l_{0})e^{-2|p_{n}|}+1}.\label{eq:Tnreg}
\end{equation}
$T_{n}^{reg}$ decreases exponentially with $n$ and, due to the translational
invariance, it is independent of $\zeta$. In the limit of $\zeta\rightarrow0$,
the sum of the first terms in $T_{n}(\zeta,0)$ was calculated in
\cite{moore2004}
\begin{equation}
\sum_{n=n_{0}}^{\infty}\frac{e^{-2|p_{n}|\zeta}}{|p_{n}|}\sim\frac{1}{\lambda\zeta}+\frac{1}{\sqrt{\lambda}}\zeta\left(\frac{1}{2},n_{0}-\frac{\varepsilon}{\lambda}\right)
\end{equation}
in terms of the Hurwitz zeta function (see \cite{DLMF}). The first,
proportional to $\zeta^{-1}$, term here is removed by the derivative
in \eqref{eq:linsys}. Then we get Eqs. \eqref{eq:linsysreg} and
\eqref{eq:Sspspp}.

For T-invariant models, when $A=0$, we have real $T_{n}(\zeta,0)$.
In the case of PBC, we can just set $l_{0}=0$ in Eqs. \eqref{eq:Tn},
\eqref{eq:Tnc}, and \eqref{eq:Tnreg} and get
\begin{align}
T_{n}(\zeta,0) & =\frac{\cos p_{n}(1-2\zeta)}{p_{n}\sin p_{n}}\quad(\lambda n<\varepsilon)\nonumber \\
T_{n}(\zeta,0) & =-\frac{1}{|p_{n}|}\frac{e^{-2|p_{n}|\zeta}+e^{-2|p_{n}|(1-\zeta)}}{1-e^{-2|p_{n}|}}\quad(\lambda n>\varepsilon,\zeta>0)\\
T_{n}^{reg} & =-\frac{2}{|p_{n}|}\frac{e^{-2|p_{n}|}}{1-e^{-2|p_{n}|}}.\nonumber 
\end{align}

In the case of HWB, substitution of Eqs. \eqref{eq:zlhwb} and \eqref{eq:enlhwb}
to \eqref{eq:Tndef} leads to
\begin{equation}
T_{n}(\zeta,\zeta')=\sum_{l=1}^{\infty}\frac{\cos(\pi l(\zeta-\zeta'))-\cos(\pi l(\zeta+\zeta'))}{\varepsilon-\lambda n-\pi^{2}l^{2}/4}.
\end{equation}
Unlike \eqref{eq:TnPBC}, it is not a function of $z-z'$ only, since
HWB is not translational invariant. Using partial fraction decomposition
and the real part of the summation formula \eqref{eq:summlexp}, we
get for $\zeta>\zeta'$
\begin{equation}
T_{n}(\zeta,\zeta')=-2\frac{\sin2p_{n}(1-\zeta)\sin2p_{n}\zeta'}{p_{n}\sin2p_{n}}.
\end{equation}
For $\lambda n>\varepsilon_{\alpha}$ and $\zeta>\zeta'$ we have
\begin{equation}
T_{n}(\zeta,\zeta')=-\frac{1}{|p_{n}|}\frac{e^{-2|p_{n}|(2-\zeta+\zeta')}+e^{-2|p_{n}|(\zeta-\zeta')}-e^{-2|p_{n}|(2-\zeta-\zeta')}-e^{-2|p_{n}|(\zeta+\zeta')}}{1-e^{-4|p_{n}|}}.
\end{equation}
The term causing the divergence is separated in the same way as in
Eq. \eqref{eq:Tnreg}, providing
\begin{equation}
T_{n}^{reg}(\zeta)=-\frac{1}{|p_{n}|}\frac{2e^{-4|p_{n}|}-e^{-4|p_{n}|\zeta}-e^{-4|p_{n}|(1-\zeta)}}{1-e^{-4|p_{n}|}}.
\end{equation}

\section{Eigenvalues of the system \eqref{eq:linsysreg} matrix\label{sec:Eigenvalues}}

Let us arrange the eigenenergies of the integrable system in increasing
order and label them by an index $k$ such that $\varepsilon_{k}\equiv\varepsilon_{n_{k}l_{k}}$
and $\varepsilon_{k}<\varepsilon_{k+1}$. The term $T_{n_{k}}(\zeta,\zeta')$
has a singularity as a function of $\varepsilon$ when $\varepsilon\rightarrow\varepsilon_{k}$
and can be separated to singular and continuous parts, $T_{n_{k}}(\zeta,\zeta')=T_{k}^{sing}(\zeta,\zeta')+T_{k}^{cont}(\zeta,\zeta')$.
For PBC, $p_{n_{k}}\sim\pi|l_{k}-l_{0}|+(\varepsilon-\varepsilon_{k})/(2\pi|l_{k}-l_{0}|)$
in the limit $\varepsilon\rightarrow\varepsilon_{k}$ and these parts
are expressed as
\begin{align}
T_{k}^{sing}(\zeta,\zeta') & =\frac{1}{2p_{n_{k}}\sin(p_{n_{k}}-\pi|l_{k}-l_{0}|)}\exp\left(2i(\pi l_{0}+\tilde{p}_{k})(\zeta-\zeta')\right)\nonumber \\
\\
T_{k}^{cont}(\zeta,0) & =-\frac{1}{2\tilde{p}_{k}}e^{2i\pi l_{0}\zeta}\left[e^{2i\tilde{p}_{k}\zeta}\left(\tan\frac{\pi(l_{k}-l_{0})+\tilde{p}_{k}}{2}+i\right)+e^{-2i\tilde{p}_{k}\zeta}\left(\cot(\pi l_{0}-\tilde{p}_{k})-i\right)\right],\nonumber 
\end{align}
where $\tilde{p}_{k}=p_{n_{k}}\mathrm{sign}(l_{k}-l_{0})$. In the
T-invariant case, when $l_{k}\neq0$, they can be expressed as 
\begin{align}
T_{k}^{sing}(\zeta,\zeta') & =\frac{\cot p_{n_{k}}}{p_{n_{k}}}\left(\cos2p_{n_{k}}\zeta\cos2p_{n_{k}}\zeta'+\sin2p_{n_{k}}\zeta\sin2p_{n_{k}}\zeta'\right)\nonumber \\
\\
T_{k}^{cont}(\zeta,\zeta') & =\frac{\sin2p_{n_{k}}(\zeta-\zeta')}{p_{n_{k}}}.\nonumber 
\end{align}
If $l_{k}=0$, $p_{n_{k}}=\sqrt{\varepsilon-\varepsilon_{k}}$, and
the second term in the parenthesis in $T_{k}^{sing}$ becomes non-singular
and is moved to $T_{k}^{cont}$. 

For HWB, when $l_{k}\neq0$, we have $p_{n_{k}}\sim\pi l_{k}/2+(\varepsilon-\varepsilon_{k})/(\pi l_{k})$
and
\begin{align}
T_{k}^{sing}(\zeta,\zeta') & =2\frac{\cot2p_{n_{k}}}{p_{n_{k}}}\sin2p_{n_{k}}\zeta\sin2p_{n_{k}}\zeta'\nonumber \\
\\
T_{k}^{cont}(\zeta,\zeta') & =-2\frac{\cos2p_{n_{k}}\zeta\sin2p_{n_{k}}\zeta'}{p_{n_{k}}}.\nonumber 
\end{align}
If $l_{k}=0$, $T_{n_{k}}(\zeta,\zeta')$ is non-singular.

\begin{figure}[H]
\includegraphics[width=8cm]{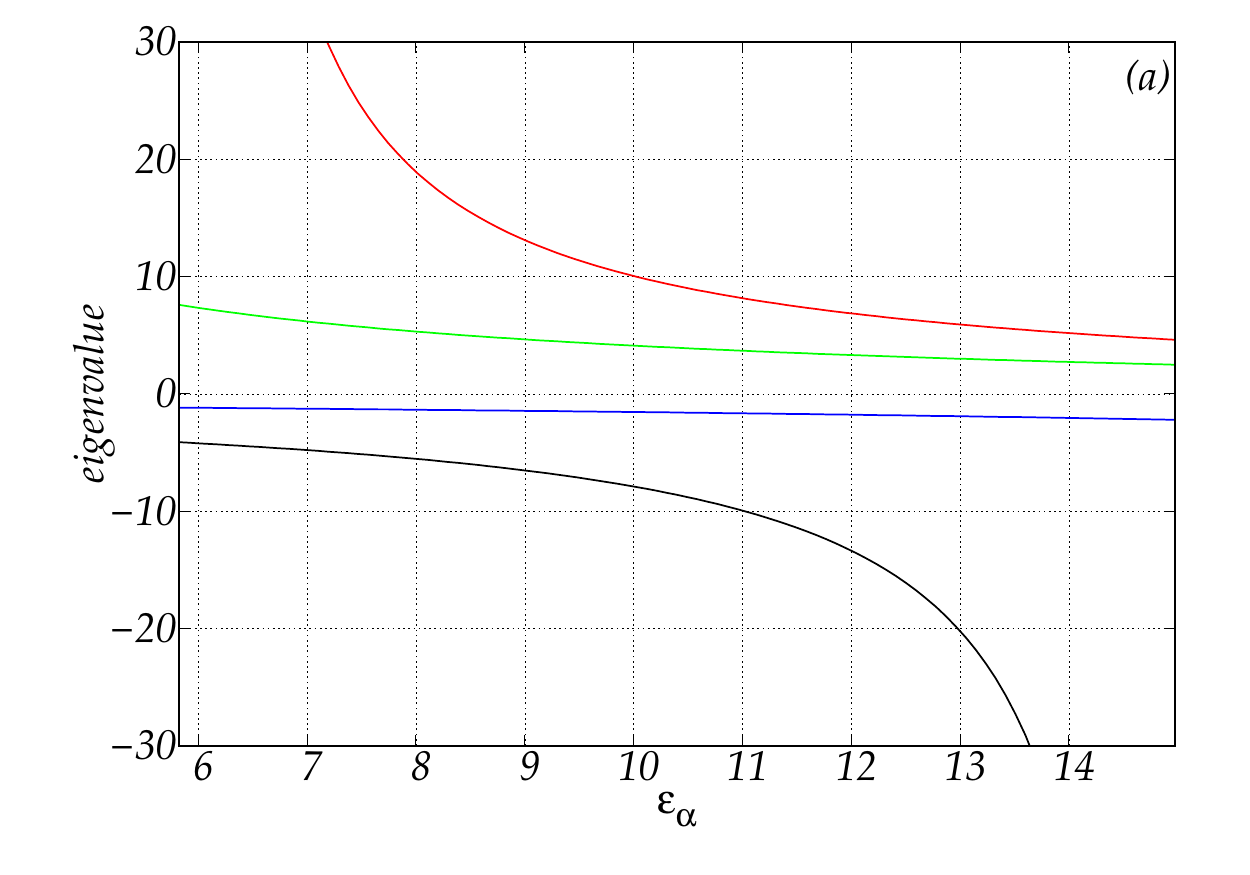}\includegraphics[width=8cm]{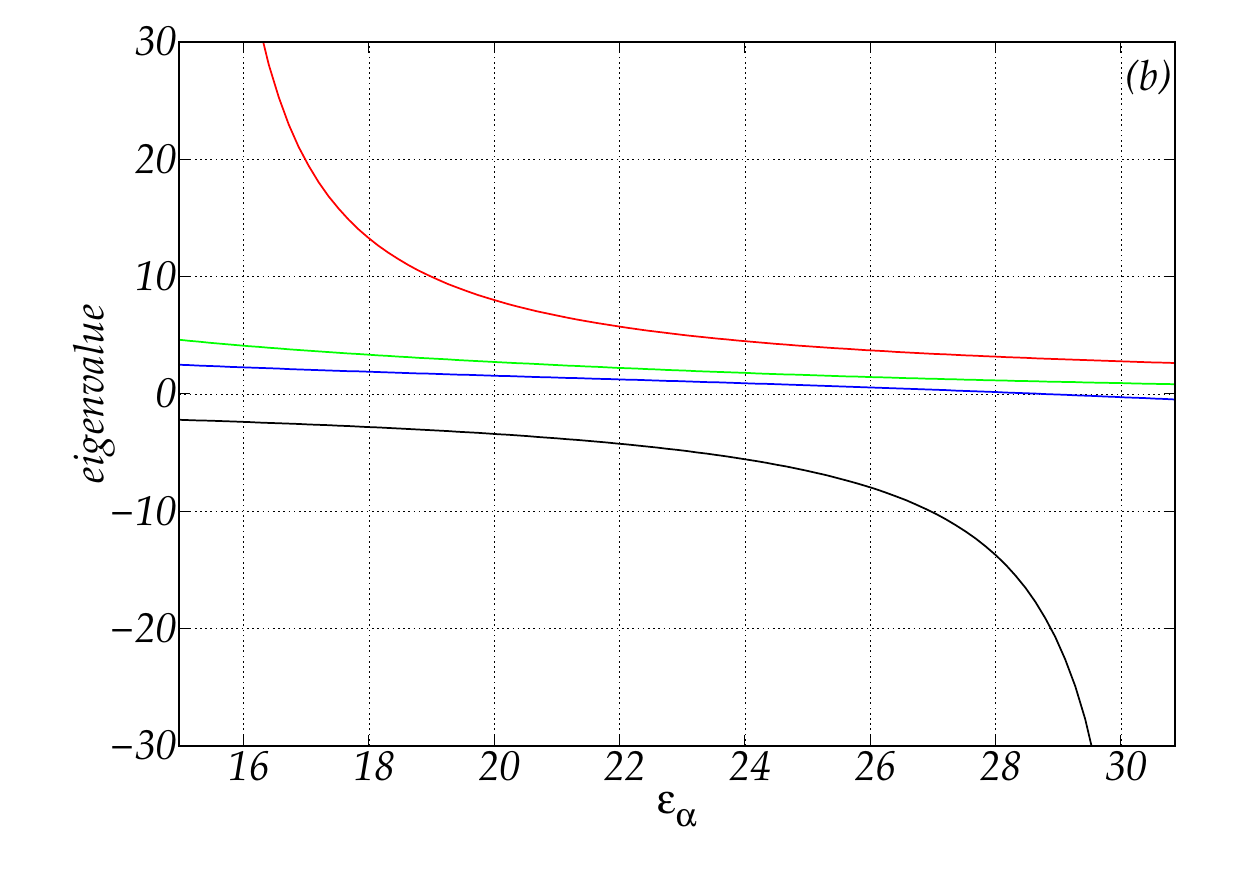}

\includegraphics[width=8cm]{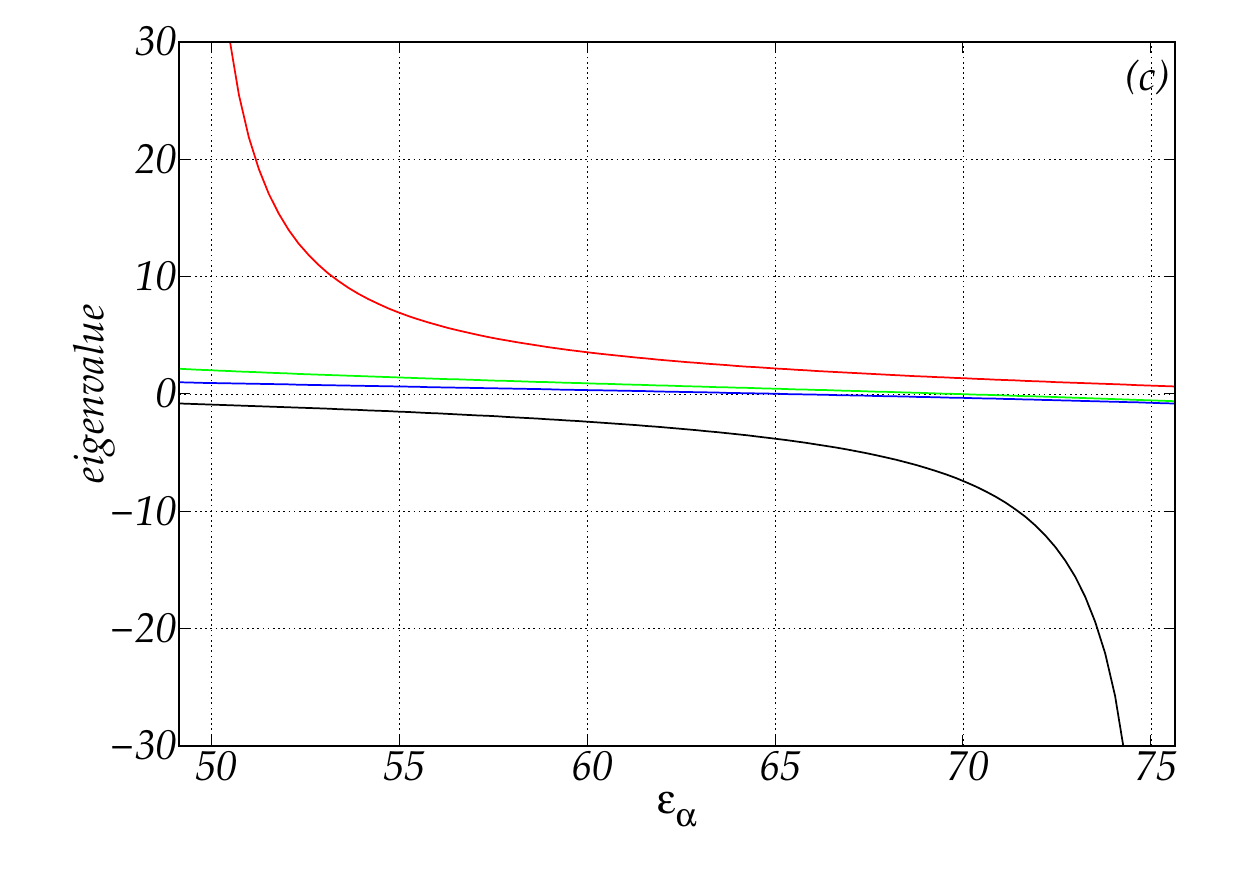}\includegraphics[width=8cm]{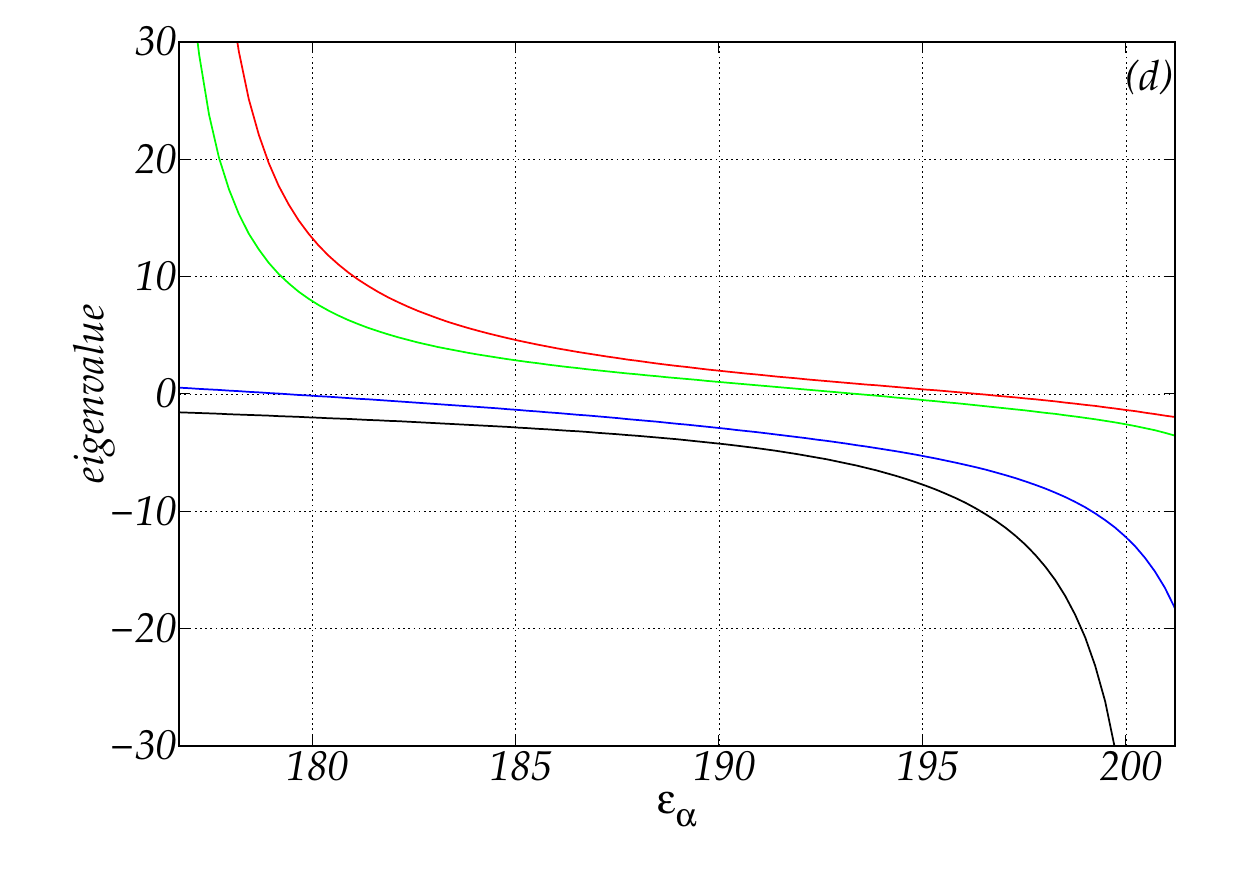}

\includegraphics[width=8cm]{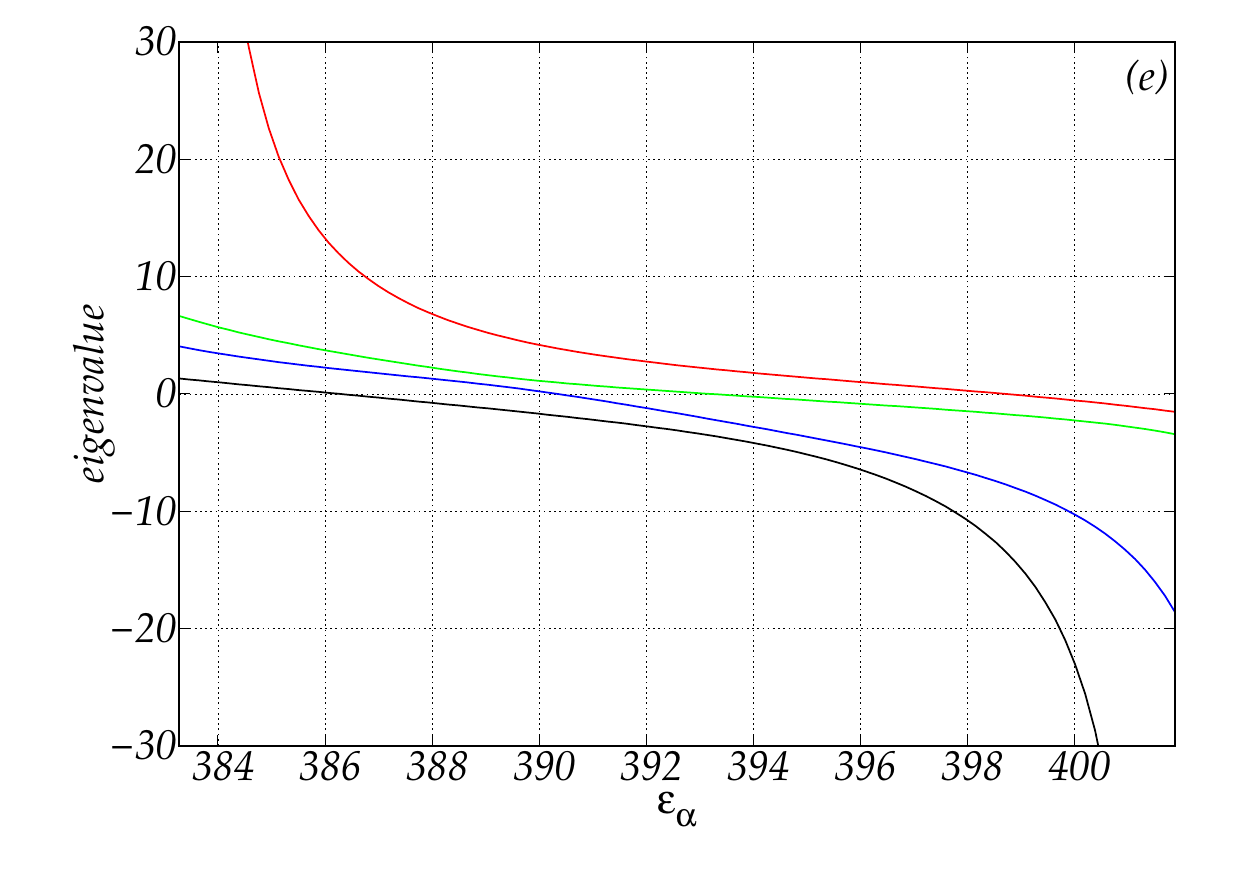}\includegraphics[width=8cm]{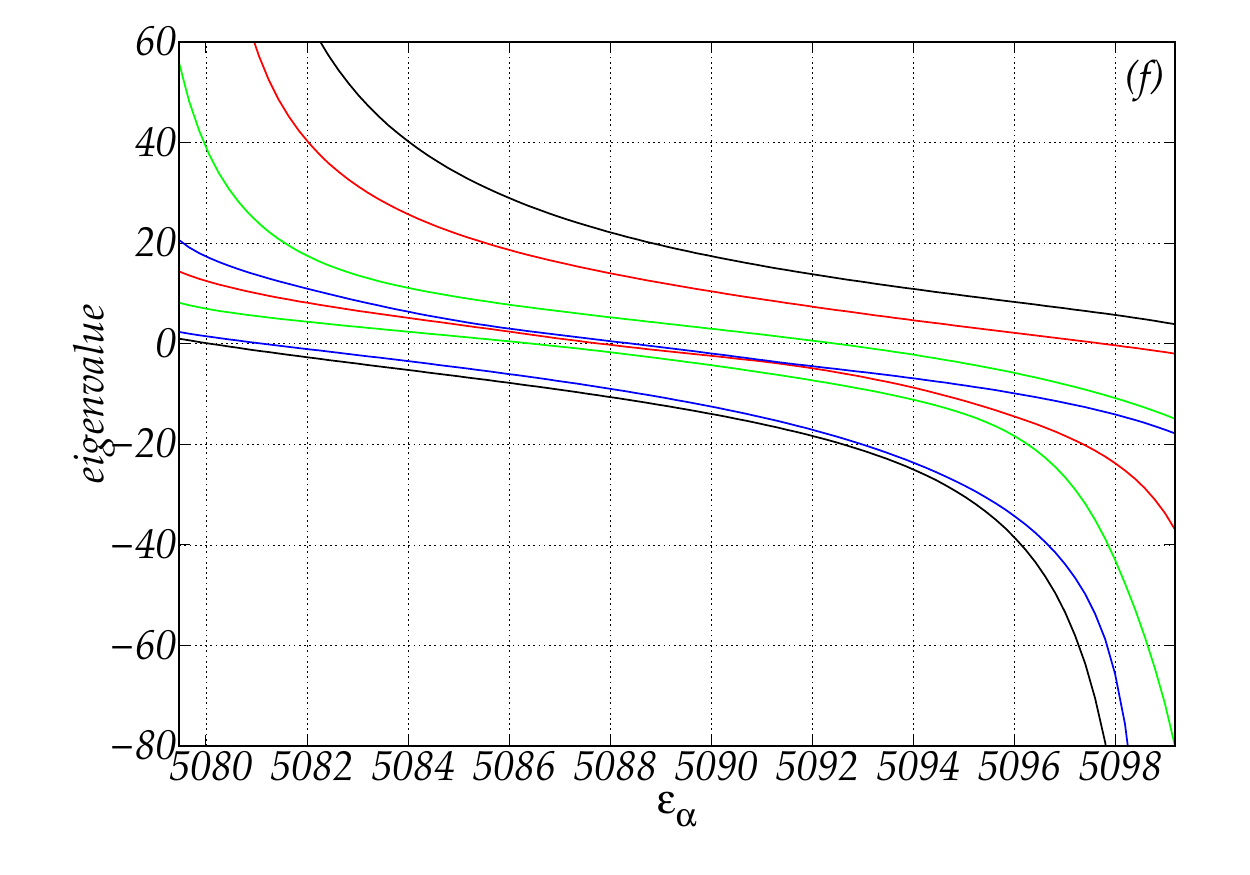}

\caption{Examples of eigenvalue dependence on the energy between two neighboring
eigenenergies of the integrable system for the non-symmetric model
with 4 scatterers (a-e) and the T-invariant model with 8 scatterers
(f). \label{fig:eigenvalue}}

\end{figure}

In any case, for $(\varepsilon_{k-1}+\varepsilon_{k})/2<\varepsilon<(\varepsilon_{k}+\varepsilon_{k+1})/2$
the matrix $S_{s's''}(\varepsilon)$ \eqref{eq:Sspspp} can be represented
as $S_{s's''}(\varepsilon)=T_{k}^{sing}(\zeta_{s'},\zeta_{s''})+S_{s's''}^{cont}(\varepsilon)$,
where $S_{s's''}^{cont}(\varepsilon)$ is continuous. The singular
part can be expressed in terms of orthonormal vectors $b_{i}(\zeta_{s'})$
\begin{equation}
T_{k}^{sing}(\zeta_{s'},\zeta_{s''})=\sum_{i=1}^{i_{max}}B_{i}b_{i}^{*}(\zeta_{s'})b_{i}(\zeta_{s''}),\quad\sum_{s'=1}^{s}b_{i'}^{*}(\zeta_{s'})b_{i}(\zeta_{s'})=\delta_{ii'}
\end{equation}
and has a form of the matrix with $i_{max}$ eigenvalues $B_{i}$.
When $\varepsilon$ approaches $\varepsilon_{k}$, the singular part
dominates and the eigenvalues tend to $\pm\infty$. Then in the T-invariant
PBC case with $l_{k}\neq0$ we have $i_{max}=2$ and two eigenvalues
of the matrix $S_{s's''}(\varepsilon)$ have singularities at $\varepsilon\rightarrow\varepsilon_{k}$,
there are no singular eigenvalues ($i_{max}=0$) in the case of HWB
with $l_{k}=0$, and single eigenvalue has a singularity in other
cases when $i_{max}=1$. Results of numerical calculations in Fig.
\ref{fig:eigenvalue} demonstrate these properties. They also show
that the eigenvalues decrease monotonically with $\varepsilon$. Then
each eigenvalue can have single root in the interval $[\varepsilon_{k},\varepsilon_{k+1}]$.
In Fig. \ref{fig:eigenvalue}, the number of eigenvalues with roots
increases from 0 to 4 in parts (a)-(e). 

In the close vicinity of $\varepsilon_{k}$ direct numerical diagonalization
of the matrix $S_{s's''}(\varepsilon)$ becomes inaccurate if $i_{max}>0$.
However, in this vicinity $i_{max}$ eigenvalues are approximated
by $B_{i}$ with good accuracy. In order to calculate other eigenvalues,
the matrix $S_{s's''}^{cont}(\varepsilon)$ is projected out of the
envelope of the vectors $b_{i}(\zeta_{s'})$, 
\begin{equation}
\sum_{s'',s'''}\left(\delta_{s's''}-\sum_{i=1}^{i_{max}}b_{i}^{*}(\zeta_{s'})b_{i}(\zeta_{s''})\right)S_{s''s'''}^{cont}(\varepsilon)\left(\delta_{s'''s^{iv}}-\sum_{i=1}^{i_{max}}b_{i}^{*}(\zeta_{s'''})b_{i}(\zeta_{s^{iv}})\right).
\end{equation}
Numerical diagonalization of this matrix provides $i_{max}$ eigenvalues
which are close to zero (they correspond to eigenvectors $b_{i}(\zeta_{s'})$),
other eigenvalues approximate the remained $s-i_{max}$ eigenvalues
of $S_{s's''}(\varepsilon)$. 

\section{Level spacing ratio\label{sec:Level-spacing-ratio}}

\begin{figure}[H]

\includegraphics[width=8cm]{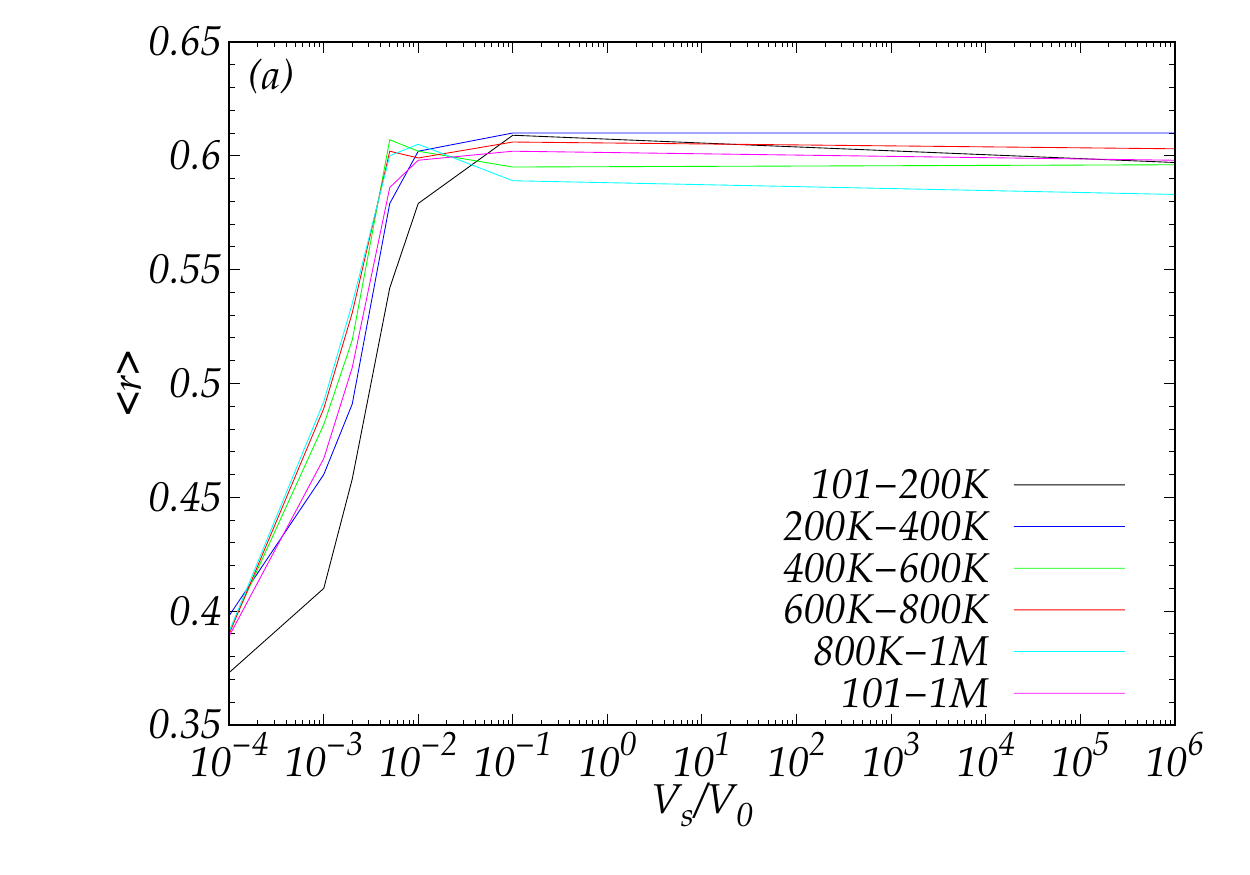}\includegraphics[width=8cm]{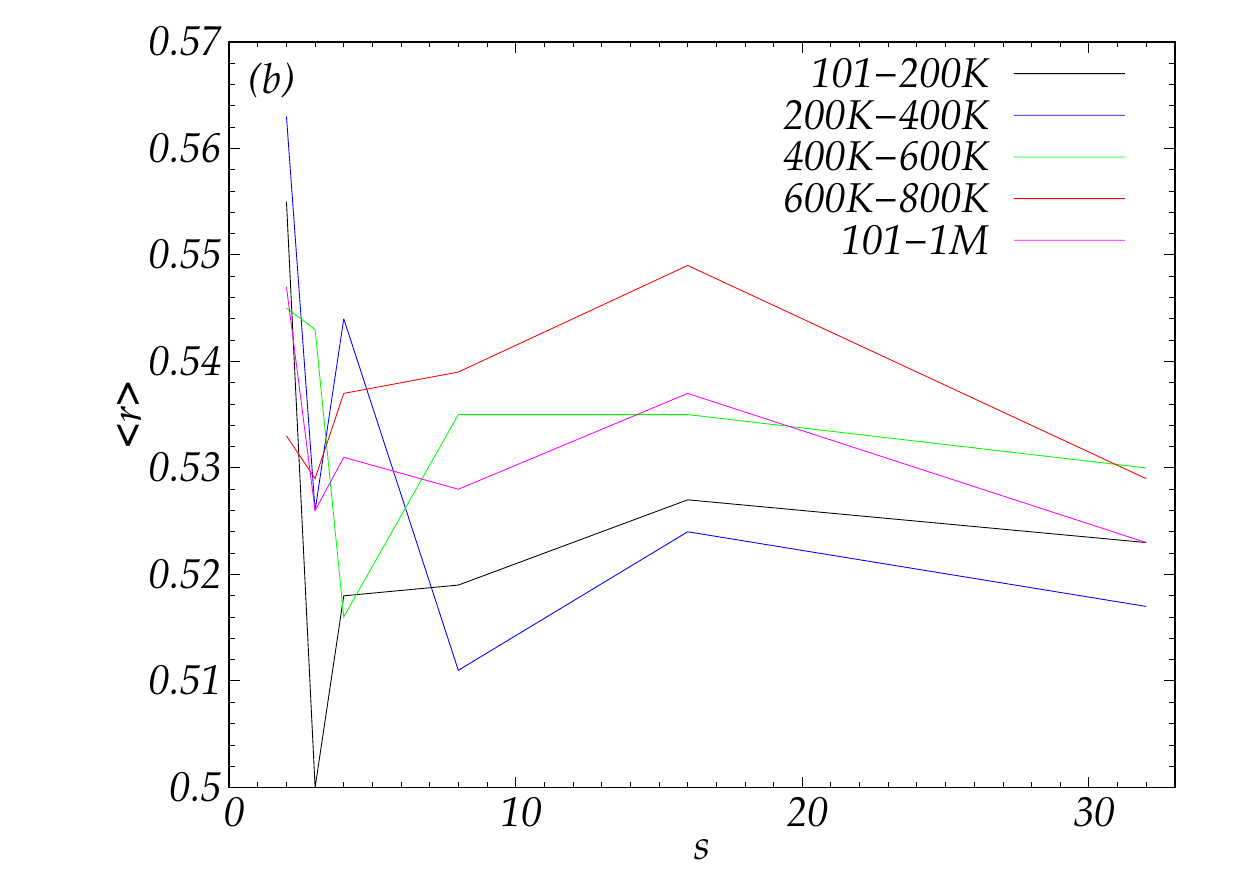}\caption{The level spacing ratio averaged over different eigenstate label intervals
(a) for the non-symmetric model as a function of the interaction strength
and (b) for the symmetric model as o function of the number of scatteres
\label{fig:LevSpRat}}

\end{figure}

The ratio of two consecutive level spacings \cite{oganesyan2007,atas2013}
\begin{equation}
r_{\alpha}=\frac{\min(E_{\alpha+1}-E_{\alpha},E_{\alpha}-E_{\alpha-1})}{\max(E_{\alpha+1}-E_{\alpha},E_{\alpha}-E_{\alpha-1})}
\end{equation}
can characterize the energy spectrum statistics and does not require
unfolding. Its averages $\left\langle r\right\rangle $ were calculated
in \cite{atas2013} for the Poisson ($\left\langle r\right\rangle =2\ln2-1\approx0.38629$),
GOE ($\left\langle r\right\rangle =4-2\sqrt{3}\approx0.53590$), and
GUE ($\left\langle r\right\rangle =2\sqrt{3}/\pi-1/2\approx0.60266$)
statistics. Figure \ref{fig:LevSpRat}(a) shows that for the present
model $\left\langle r\right\rangle $ increases at weak interactions,
but demonstrate non-monotonic dependence when the value $\left\langle r\right\rangle \approx6$,
corresponding to GUE, is approached. In some eigenstate intervals,
the level spacing ratio has maximum $\left\langle r\right\rangle \approx6$
already at $V=5\times10^{-3}V_{0}$, in contradiction with NND and
spectral rigidity (cf. Fig. \ref{fig:NNDSpRigV}). When the number
of scatterers is increased (see Fig. \ref{fig:LevSpRat}(b)), $\left\langle r\right\rangle $
non-monotonically decreases, although the monotonic increase of chaoticity
is demonstrated by the NND change from \v{S}eba to GOE predictions,
as well as by the spectral rigidity (see Fig. \ref{fig:NNDSpRigSym}). 

\section{Expectation values\label{sec:Expectation-values}}

Substituting Eqs. \eqref{eq:nlalpha} and \eqref{eq:ExpValPotn} into
Eq. \eqref{eq:ExpValalphan} we can get the following expression for
the expectation value of the transverse potential energy in the non-integrable
system eigenstates
\begin{multline}
\left\langle \alpha\left|\frac{1}{2}m\omega_{\perp}^{2}\rho^{2}\right|\alpha\right\rangle =\frac{\omega_{\perp}}{2}\mathcal{N}_{\alpha}\sum_{l=-\infty}^{\infty}\Lambda_{l}\sum_{n,n'}\frac{\left(2n+1\right)\delta_{nn'}-n\delta_{n'n-1}-n'\delta_{nn'-1}}{(\varepsilon_{\alpha}-\varepsilon_{nl})(\varepsilon_{\alpha}-\varepsilon_{n'l})}.\\
=\frac{\omega_{\perp}}{2}\left\{ 1+2\frac{\mathcal{N}_{\alpha}}{\lambda^{2}}\sum_{l=-\infty}^{\infty}\Lambda_{l}\sum_{n=0}^{\infty}\left[\frac{n}{(q_{l}+n)^{2}}-\frac{n}{(q_{l}+n)(q_{l}+n-1)}\right]\right\} ,\label{eq:ExpValPotalpha}
\end{multline}
where the last transformation uses the normalization condition, Eq.
\eqref{eq:Enl} for $\varepsilon_{nl}$, and Eq. \eqref{eq:ql} for
$q_{i}$. The sum over $n$ here can be transformed as
\begin{equation}
\sum_{n=0}^{\infty}\left[-\frac{q_{l}}{(q_{l}+n)^{2}}+\frac{q_{l}-1}{(q_{l}+n)(q_{l}+n-1)}\right]=-q_{l}\zeta(2,q_{l})+(q_{l}-1)\sum_{n=0}^{\infty}\left(\frac{1}{q_{l}+n-1}-\frac{1}{q_{l}+n}\right),
\end{equation}
where the summation over $n$ with Eq. \eqref{eq:Sumnzeta} for the
first term in the square brackets and partial fraction decomposition
for the second term are used. The last sum over $n$ is reduced to
$1/(q_{l}-1)$ due to cancellation of the terms. This leads to Eq.
\eqref{eq:ExpValUperp}. The derivation above is related to the T-noninvariant
models. The same transformation of the sum over $n$ leads to the
expectation values for the T-invariant PBC \eqref{eq:ExpValUperpT}
and HWB \eqref{eq:ExpValUperpB} models.

The variance between the integrable system eigenstates can be evaluated
analytically. The product $\bar{\alpha}(\varepsilon)\overline{U_{\perp}}$can
be approximated by the sum
\begin{equation}
\sum_{n,l}\left\langle nl\left|\hat{U}_{\perp}\right|nl\right\rangle \theta(\varepsilon-\varepsilon_{nl})=\frac{\lambda}{2}\sum_{n=0}^{[\varepsilon/\lambda]}(n+\frac{1}{2})\sum_{l=l_{min}(n)}^{l_{max}(n)}\frac{1}{\varepsilon_{nl}+\lambda/2},
\end{equation}
where $l_{min,max}(n)=l_{0}\mp\sqrt{\varepsilon-\lambda n}/\pi$ and
Eqs. \eqref{eq:Enl} and \eqref{eq:ExpValPotn} are used. Replacing
summation by integration and neglecting the values $\sim1$ compared
to $n$, we approximate the sum as
\begin{equation}
\intop_{0}^{\varepsilon/\lambda}ndn\intop_{l_{min}}^{l_{max}}dl\frac{1}{\varepsilon_{nl}}=\frac{4}{9\pi\lambda}\varepsilon^{3/2}.
\end{equation}
It has the same $\varepsilon$ dependence as $\bar{\alpha}(\varepsilon)$
taken with the same accuracy {[}the first term in Eq. \eqref{eq:alphabarPBC}{]}.
Then the average $\overline{U_{\perp}}=1/3$ is independent of the
averaging interval (this value agrees to the virial theorem). In the
same way we find $\overline{U_{\perp}^{2}}=2/15$ and, therefore,
$\mathrm{Var}_{nl}(\hat{U}_{\perp})=1/45$. Although in the HWB model
$l\geq0$, we get the same results due to the distinction between
Eqs. \eqref{eq:Enl} and \eqref{eq:enlhwb}.

For the average axial momentum, we approximately evaluate the sum
\begin{equation}
\sum_{n,l}l\theta(\varepsilon-\varepsilon_{nl})\approx\intop_{l_{min}(0)}^{l_{max}(0)}ldl\intop_{0}^{n_{max}}dn=\frac{4}{3\pi\lambda}\varepsilon^{3/2}l_{0},
\end{equation}
where $n_{max}=[\varepsilon-\pi^{2}(l-l_{0})^{2}]/\lambda$. This
leads to $\overline{p_{ax}}=l_{0}$. However, evaluating $\overline{p_{ax}^{2}}$,
we see that
\begin{equation}
\sum_{n,l}l^{2}\theta(\varepsilon-\varepsilon_{nl})\approx\frac{4}{3\pi\lambda}\left(\varepsilon^{3/2}l_{0}^{2}+\frac{1}{5\pi^{2}}\varepsilon^{5/2}\right)
\end{equation}
has a different $\varepsilon$ dependence. Therefore, 
\begin{equation}
\overline{p_{ax}^{2}}=l_{0}^{2}+\frac{1}{5\pi^{2}}\frac{\varepsilon_{max}^{5/2}-\varepsilon_{min}^{5/2}}{\varepsilon_{max}^{3/2}-\varepsilon_{min}^{3/2}}
\end{equation}
depends on the averaging interval $[\varepsilon_{min},\varepsilon_{max}]$
boundaries. As a result, we get the variance \eqref{eq:Varnlpax}.

In the integrable system basis, $\overline{P_{pos}}=\overline{P_{odd}}=\overline{P_{pos}^{2}}=\overline{P_{odd}^{2}}=1/2$.
This leads to $\mathrm{Var}_{nl}(\hat{P}_{pos})=\mathrm{Var}_{nl}(\hat{P}_{odd})=1/4$.




\nolinenumbers

\def\cited#1{{\hspace{\stretch{1}} \mbox{\textit{Cited:} \textbf{#1}}}}

\end{document}